%% file: main.tex
\documentclass[authorversion,sigconf,nonacm]{acmart}

\AtBeginDocument{%
  \providecommand\BibTeX{{%
    \normalfont B\kern-0.5em{\scshape i\kern-0.25em b}\kern-0.8em\TeX}}}

\copyrightyear{2024}
\acmYear{2024}
\setcopyright{acmlicensed}
\acmConference[CHI '24]{Proceedings of the CHI Conference on Human Factors in Computing Systems}{May 11--16, 2024}{Honolulu, HI, USA}
\acmBooktitle{Proceedings of the CHI Conference on Human Factors in Computing Systems (CHI '24), May 11--16, 2024, Honolulu, HI, USA} \acmDOI{10.1145/3613904.3642852} \acmISBN{979-8-4007-0330-0/24/05}

\usepackage{_macros}
\usepackage{tikz}
\newcommand*\circled[1]{\tikz[baseline=(char.base)]{
            \node[shape=circle,draw,inner sep=0.5pt] (char) {#1};}}
\usepackage{multirow}
\usepackage{caption}
\usepackage{subcaption}
\usepackage{array}
\usepackage{geometry}
\usepackage{multirow}
\usepackage{array}
\usepackage{enumitem}
\usepackage{caption}
\usepackage{lipsum}
\usepackage{ulem}

\begin{document}

\title{"When He Feels Cold, He Goes to the Seahorse"—Blending Generative AI into Multimaterial Storymaking for Family Expressive Arts Therapy}
\titlenote{To appear at ACM CHI '24}


\author{Di Liu}
\orcid{}
\affiliation{%
  \institution{School of Design, SUSTech}
  \city{Shenzhen}
  \country{China}
}
\email{seucliudi@gmail.com}

\author{Hanqing Zhou}
\affiliation{%
  \institution{School of Design, SUSTech}
  \city{Shenzhen}
  \country{China}
}
\email{12331483@mail.sustech.edu.cn}

\author{Pengcheng An}
\authornote{Corresponding Author.}
\affiliation{%
  \institution{School of Design, SUSTech}
  \city{Shenzhen}
  \country{China}
}
\email{anpc@sustech.edu.cn}

\renewcommand{\shortauthors}{D. Liu, H. Zhou \& P. An}

\begin{abstract}
Storymaking, as an integrative form of expressive arts therapy, is an effective means to foster family communication. Yet, the integration of generative AI as expressive materials in therapeutic storymaking remains underexplored. And there is a lack of HCI implications on how to support families and therapists in this context. Addressing this, our study involved five weeks of storymaking sessions with seven families guided by a professional therapist. In these sessions, the families used both traditional art-making materials and image-based generative AI to create and evolve their family stories. Via the rich empirical data and commentaries from four expert therapists, we contextualize how families creatively melded AI and traditional expressive materials to externalize their ideas and feelings. Through the lens of Expressive Therapies Continuum (ETC), we characterize the therapeutic implications of AI as expressive materials. Desirable interaction qualities to support children, parents, and therapists are distilled for future HCI research.
\end{abstract}



\begin{teaserfigure}
  \includegraphics[width=\textwidth]{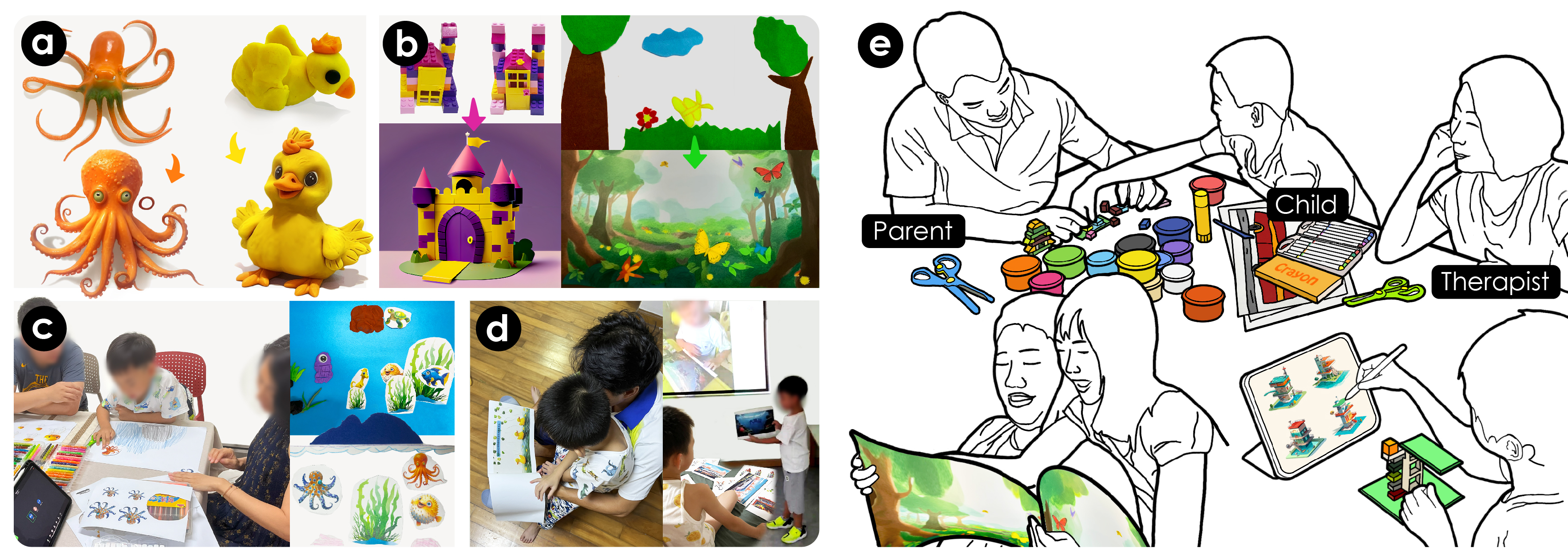}
  \vspace{-7mm}
  \caption{(a)~Character Making; (b)~Scene/Object Making; (c)~Story Making and Storytelling; (d)~Story Sharing and Reflection; (e)~Setup Display of Family Expressive Arts Therapy Activity.}
  \Description{In Figure a, an orange octopus toy is transformed into a cartoon character octopus image, and a yellow chick created using clay is transformed into a cute yellow chick image. In Figure b, a colorful castle built using different colors of Lego is displayed. Additionally, a forest scene created using brightly colored cloth collage is transformed into a beautiful forest. Figure c illustrates a father and his son telling stories together using AI-infused multi-material. In Figure d, the father and his son are shown at home sharing the creation of physical storybooks, and children are seen sharing their own storybooks with others in kindergarten. Figure e uses simple drawings to depict the layout of activity scenes.Parents and children collaborate using physical materials to create artifacts together, while the therapist observes their creative process. When family-created artifacts and verbal descriptions of the family are fed into the generative AI as prompts, children and parents can choose to discuss them together. The final storybook is taken home and shared with parents.}
  \label{fig:teaser}
\end{teaserfigure}


\maketitle

\input{texfiles/1-introduction}

\input{texfiles/2-relatedwork}

\input{texfiles/3-Pilot_Study}

\input{texfiles/4-Methods}

\input{texfiles/Findings}

\input{texfiles/7-discussion}

\input{texfiles/8-conclusion}

\bibliographystyle{ACM-Reference-Format}
\bibliography{reference.bib}

\end{document}

%% file: texfiles/1-introduction.tex
\section{Introduction}
Family communication plays a pivotal role in co-constructing and negotiating meanings, shaping identity, and influencing the dynamics of parent-child relationships~\cite{braithwaite2017engaging}. It can also support children's psychological well-being and overall development~\cite{kavehfarsani2020study}.
However, family communications can be hindered or disrupted by many issues, such as excessive technology use~\cite{chin2022s}, parental work–family conflict~\cite{morr2020parent}, and undesirable communication patterns~\cite{ritchie1990family}. 
This could result in significant and lasting negative effects, including increased family tension or conflict~\cite{shin2021designing}, mental health issues of children~\cite{barrett2001cognitive}, and behavioral problems and at-risk behaviors of the family unit~\cite{xiao2011perceptions}. 
Hence, it is essential to enhance the understanding of family dynamics and foster effective family communication.
    
Family expressive arts therapy, facilitated by expressive art therapists, 
engages family groups in integrative and multi-modal expressive and constructive tasks—e.g., drawing, playing, storytelling, or collage-making—which can help break down barriers of communication and develop a collaborative approach to explore family dynamics and strengthen family relationships~\cite{malchiodi2013expressive,kwiatkowska1967family,levick1973family}.
Storymaking has been one of such therapeutic activities frequently leveraged by therapists with family groups, especially children and parents~\cite{friedland2010stories,millerstory}. 
In a storymaking-based expressive arts therapy session, therapists usually guide family members to co-create therapeutic stories that externalize their inner thoughts, reflect their shared experiences, and encourage mutual understanding through the integration of art-making and storytelling.
Despite these benefits for families, research in Human-Computer Interaction (HCI) has accumulated little knowledge about how to support families and therapists in storymaking-based expressive arts therapy sessions.
 
Nonetheless, outside the settings of expressive arts therapy, a substantial body of studies focus on supporting storytelling and storymaking for children and families~\cite{zhang2022storybuddy,dietz2021storycoder,rubegni2014fiabot,panjwani2017constructing}. As distinguished by prior research, storytelling focuses on the act of narrating a story, whereas storymaking refers to creating a story by integrating the inventive modes of storytelling and crafting~\cite{panjwani2017constructing}.
Much research in HCI has explored how to support storytelling to facilitate various purposes, such as remote communication~\cite{raffle2010family,raffle2011hello}, language learning~\cite{kory2014storytelling,druin2009designing}, physical fitness~\cite{saksono2017reflective}, and mathematics learning~\cite{alexandre2010maths4kids}, to name but a few. 
On the other hand, many studies also encompassed children's storymaking: creating physical or digital stories to, for instance, promote children's creativity~\cite{zarei2021towards,lu2011shadowstory}, emotional expression learning~\cite{schlauch2022investigating,ryokai2012storyfaces}, or computational thinking and programming~\cite{dietz2021storycoder,dietz2023visual}. Among them, relatively a small number of studies concerned how to support children and their families to collaborate in crafting stories. As an example, Torben et al. developed a tangible storytelling system to aid long-distance relationships between grandparents and grandchildren through crafting~\cite{wallbaum2018supporting}. 
Ricarose et al. examined the benefits and challenges of involving youth and families in creating stories using the computational constructionkit~\cite{roque2021opportunities}.
While above mentioned research yielded important insights, relatively less work has addressed family storymaking in co-located, multilateral settings (a common setup in family expressive therapy sessions). To the best of our knowledge, there is a lack of research into the implications of supporting both the family (children and parents) and the therapists when using storymaking in the settings of expressive arts therapy. 

Materials occupy an essential presence in expressive arts therapy as they afford different expressive properties that can evoke specific experiences during the act of art-creating~\cite{lazar2018making,regev2017parent}.
While the properties of traditional physical art-making materials have been extensively examined in expressive therapy studies~\cite{kagin1978expressive,malchiodi2022handbook}, it is yet to be explored whether and how digital media, including the emerging generative AI techniques, could be utilized as materials for expressive arts therapy.
In recent years, the concept of ``AI as a design material'' has presented promising and novel prospects for designers or design-oriented research in HCI~\cite{dove2017ux,holmquist2017intelligence}. 
In particular, with the advent of image-based generative AI tools like Dall·E 2~\cite{dawebsite}, Stable Diffusion~\cite{sdwebsite}, and Midjourney~\cite{midwebsite}, some studies have delved into the exploration of integrating text-to-image models as design materials into design practices~\cite{kulkarni2023word,tholander2023design}.
These studies revealed the potential properties of generative AI as design materials to propel the creativity of (both amateur and professional) designers, suggesting an unaddressed opportunity to explore the properties of generative AI when used as expressive materials in expressive arts therapy.

Therefore, corresponding to the above-mentioned opportunities, the present study set out to address a three-fold research question:

\begin{itemize}
  \item \textbf{\textit{RQ1: how would families engage in AI-infused multimaterial storymaking in expressive arts therapy?}}
  \item \textbf{\textit{RQ2: how would image-generative AI be leveraged as expressive materials in family expressive arts therapy?}}
  \item \textbf{\textit{RQ3: how would future design support parents, children and therapists in family expressive arts therapy?}}
\end{itemize}

To this end, we conducted family storymaking based expressive arts therapy activities with a professional therapist based on an ongoing family expressive arts therapy activity. We report our findings based on five-week sessions involving 7 groups of families with 18 participants~(10 parents, 8 children) and interview data with the parents. Further, we conducted a discussion with 4 expert therapists in the form of a focus group in order to gather commentaries about our data from a more diverse therapeutic perspective. Our findings contextualize the creative process of the families and the therapeutic meaning behind it. Through the lens of Expressive Therapies Continuum (ETC), we characterize the therapeutic implications of the generative AI as expressive materials. This paper provides relevant desirable interaction qualities to support children, parents and therapists for future HCI research.

This paper contributes (i) empirical understandings on how to support story-making based expressive arts therapy using the generative AI; and (ii) design insights about desirable interaction qualities for children , parents, and therapists, and design implications for HCI systems to support AI-infused family expressive arts therapy.

%% file: texfiles/2-relatedwork.tex
\section{Background and Related Work} \label{sec:background}

\subsection{Storymaking in Expressive Arts Therapy}
Expressive arts therapy has been proven to be an effective intervention to understand family dynamics and foster family communication~\cite{kwiatkowska1967family, harvey1990dynamic,lai2011expressive}. It refers to an integrative and multi-modal intervention incorporating multiple art-making forms to achieve personal growth and social transformation~\cite{malchiodi2003expressive,levine1998foundations}. Within expressive arts therapy, the focus lies on the \textit{process of creation} rather than the \textit{product of creation}. Its benefits have been evident across people of all ages~\cite{rubin1999art,du2024deepthink}. 
In an expressive arts therapy session, an expressive arts therapist observes and encourages clients to trigger personal growth and explore the thought through creative processes of expression, including drawing, movement, drama, play, and storytelling. 

Notably, storymaking, as a form of expressive arts therapy that combines various types of creative and expressive tasks, has been commonly leveraged with families~\cite{friedland2010stories,millerstory,harpaz2016storytelling,matthews2011my,brosnan2006evaluation}. For example, Lai et al. utilized expressive arts therapy to enhance mother-child relationships between abused mothers and their children and facilitate their personal transformation~\cite{lai2011expressive}.
Such therapeutic storymaking often incorporates art-making techniques~(e.g., drawing, clay manipulation, and collage-making) with storytelling~\cite{friedland2010stories,neill2003storymaking}. 
In the process of family members' art-making, therapists can provide a comfortable and safe space where parents and children utilize physical expressive materials as languages for social, physical, and spiritual wellness~\cite{lazar2018making}; in addition, during their therapeutic narrative creating process, the characters they created serve as metaphors, projecting theirs own inner thoughts~\cite{perrow2008healing}. When developing story plots, therapists would guide participants to build stories' tension by setting obstacles to encourage participants to solve problems~\cite{perrow2008healing}. However, there is a lack of knowledge about the role of technology in family expressive arts therapy.

In storymaking-based expressive arts therapy, clients tend to interact with diverse expressive materials. Material interaction plays a pivotal role in expressive arts therapy, facilitating self-expression and therapeutic healing via creative expression~\cite{lazar2018making,hinz2019expressive}. The therapists are trained to understand the properties of various materials and their potential therapeutic impacts. The Expressive Therapies Continuum (ETC) framework offers valuable insights for addressing this aspect~\cite{hinz2019expressive}. ETC serves as a valuable tool for categorizing and organizing how clients interact with expressive materials as they process information and form images~\cite{kagin1978expressive}. It involved three horizontal levels~(\textit{Kinesthetic/Sensory}, \textit{Perceptual/Affective}, and \textit{Cognitive/Symbolic}) of information processing, which is bipolar or complementary, along with a vertical \textit{creative level} as the integration of any or all of the levels~\cite{lusebrink2015expressive}.

The current practices of expressive arts therapy mainly rely on traditional materials. Few studies introduced digital materials, and little is known about how to leverage generative AI as expressive materials to support family expressive arts therapy.

\subsection{Storytelling and Storymaking in HCI}
Storymaking as expressive arts therapy, remains scarcely studied in HCI. However, a large body of HCI research has explored technologies supporting children or families' storytelling and storymaking outside the settings of expressive arts therapy~\cite{santos2020therapist,alexandre2010maths4kids,zhang2022storydrawer,sylla2015investigating,saksono2017reflective}.
Storytelling and storymaking offer distinct avenues for creative expression, with storytelling enabling personal narrative representation and storymaking facilitating the creation of physical representations to reflect and convey our experiences and observations~\cite{panjwani2017constructing}. 
The HCI domain has witnessed a surge of interest in designing systems involving storytelling for facilitating various applications beyond entertainment, such as social engagement~\cite{sargeant2014far,raffle2010family}, development of literacy~\cite{kory2014storytelling,michaelis2017someone}, and social-emotional learning~\cite{santos2020therapist}. 
For example, a companion robot for children learning new vocabulary through a storytelling game~\cite{kory2014storytelling}. Saksono et al. explored using storytelling to encourage families to reflect on factors influencing children's physical activity~\cite{saksono2017reflective}.

On the other hand, HCI research has designed applications and systems to support children to create multimedia story content (e.g., text, audio or photo), in order to facilitate their computational thinking and programming~\cite{dietz2023visual,dietz2021storycoder}, collaboration skills~\cite{rubegni2014fiabot}, creativity~\cite{zarei2020investigating}, and gender stereotype awareness~\cite{rubegni2019detecting,rubegni2022raising}. For instance, StoryDrawer~\cite{zhang2022storydrawer} supports creativity and visual storytelling through co-drawing. Rubegni et al. presented a collaborative digital storytelling tool to help children notice the negative gender stereotypes~\cite{rubegni2022raising}. 
Moreover, a relatively smaller body of research has focused on the storymaking by family members. 
The StoryBox~\cite{wallbaum2018supporting} is a tangible system facilitating the long-distance, asynchronous sharing of co-created images, written messages, and audio between children and their grandparents.
Moreover, Roque et al. explored the opportunities and limitations arising from involving youth and their families in the creation and sharing of stories using a computational construction kit~\cite{roque2021opportunities}. Also, Yu et al. examined family dynamics during family joint storymaking, focusing on the negotiation practices among family members~\cite{yu2023family}.

An emerging stream of storymaking systems also featured the integration of physical and digital experiences for children. 
These tools have been successfully leveraged to support various aspects of learning and skills development, such as narrative development~\cite{budd2007pagecraft,sylla2015investigating}, emotional learning~\cite{cai2023emotionblock,ryokai2012storyfaces}, and fantasy play and storytelling~\cite{cassell2001making}. For example, PageCraft is an interactive storytelling that can connect physical media and digital media by transforming children's real-world play into text and visuals displayed on a screen~\cite{budd2007pagecraft}. 
However, many applications in the realm of storymaking rely on pre-established story templates or a structured framework which may inadvertently limit the scope of creativity~\cite{schlauch2022investigating}. 
Several studies set out to address this constraint by facilitating children's story creation through connecting physical and digital environments~\cite{rubegni2018design,schlauch2022investigating}. 
As shown by prior research, introducing physical materials grants children extra developmental benefits in creative process~\cite{resnick2017lifelong,kirsh2013embodied}.
For the same reason, when used in family expressive arts therapy, storymaking often relies on various physical materials. However, HCI research has sparsely explored how to support co-located multimaterial storymaking in family expressive arts therapy. And relevant understanding is yet to be established on whether and how generative AI could be integrated as a form of expressive materials in this context. 


\subsection{Child or Family Engagement with AI}
In HCI, there is a substantial body of research exploring how to design technologies to promote children's and families' AI literacy~\cite{long2023fostering,adisa2023spot,druga2022family}.
In addition, some studies have explored how to utilize AI algorithms in designed systems to ease children' daily activities, such as medical diagnosis~\cite{mohanta2020classifying}, emotion recognition~\cite{baldovino2019child} or speech recognition~\cite{wollmer2011tandem}.
Furthermore, a large number of studies have shown conversational agents and embodied robotic interfaces as a collaborative partner and playmate to facilitate children's learning and skill development, such as question answering behavior~\cite{lee2023dapie}, science or math learning~\cite{xu2022contingent}, social play~\cite{pantoja2019voice},and growth mindset~\cite{park2017growing}. 
For example, Ying et al. demonstrated that a conversational agent has the potential to boost children's engagement in narrative-relevant vocalizations, diminish unrelated vocalizations, and elevate story comprehension~\cite{xu2022dialogue}.
In particular, some of studies introduced conversational AI integrating storytelling strategies. For example, Kory et al. have shown that using a social robot narrating a story in a game could enhance children's early language learning~\cite{kory2014storytelling}. Also, StoryCoder~\cite{dietz2021storycoder} uses a voice agent to teach Computational Thinking (CT) concepts. However, while above studies focused on child-facing AI interfaces, some studies also involved families.

Families as an essential factor of children's development, also play a significant role in children’s interaction with technology~\cite{cagiltay2023family,beneteau2020parenting}. 
In contrast to the substantial body of research focused on children, relatively a smaller proportion of research explored how conversational agents could support both children and their family in, for instance, joint reading~\cite{xu2023rosita,zhang2022storybuddy}, family communication~\cite{beneteau2019communication,beneteau2020parenting}, and mathematical learning~\cite{xu2023mathkingdom}.  
Namely, Beneteau et al. explored using smart speakers at home to support parent-child communication and augment parenting practices~\cite{beneteau2020parenting}. Zhang et al. presented a human-AI storytelling system which can enhance children’s engagement in the storytelling process through parent involvement~\cite{zhang2022storybuddy}.
These studies have yielded valuable insights into the design of conversational AI to support children and family.

Besides conversational AI, the image generative AI models, such as Recurrent Neural Network~(RNN)~\cite{grossberg2013recurrent} or Generative Adversarial Networks~(GAN)~\cite{goodfellow2014generative}, has also initiated its exploration into supporting children's creative activities. 
Many studies have explored how RNN-based models can be used for child-AI co-sketching to support children's creativity~\cite{zhang2022storydrawer,ali2020can,ali2021social,zhang2023observe}. For example, Magic Draw serves as a collaborative drawing game to cultivate children's figurative creativity through verbal interactions between children, a robot, and a tablet~\cite{ali2020can}. Also, Bio Sketchbook, an interactive tool, helps children observe and document biodiversity, enhancing their nature connection and observation skills~\cite{zhang2023observe}.
Furthermore, with the recent development of Diffusion models~\cite{ho2020denoising}, it can present a promising opportunity for the expansion and innovation of image applications. A recent study has demonstrated the efficacy of diffusion model-based image generation tools in facilitating personalized and adaptive visual storytelling tailored for children~\cite{han2023design}.
However, prior work has rarely explored on how Diffusion models can be employed as family co-creative support. Further, few studies focus on how Diffusion models incorporate into family expressive arts therapy.

%% file: texfiles/3-Pilot_Study.tex
\section{Context Study}
As a preparation for the later stage, we conducted a context study using semi-structured online interviews with two professional expressive arts therapists. The context study served two purposes:
first, to concretely understand the current practice of family expressive arts therapy; second, to discuss whether image-based generative AI could potentially support families and therapists in the context.
Both therapists came with over 5 years of expertise in practice. In the interviews, they detailed their existing approach to family expressive arts therapy sessions. Following that, they were presented with several image-based generative AI tools and were asked for their thoughts on whether and how these tools could be used in their practice.

As revealed by the therapists, families typically enroll in family expressive arts therapy because parents seek a deeper understanding of their children, particularly the facets not readily apparent to them. They also aim to discover improved communication methods that can bolster their child's growth and maintain strong parent-child bonds. Contrary to some assumptions, family expressive arts therapy is not exclusive to families grappling with significant relationship challenges or mental health concerns of members. In fact, a majority of their clientele consists of typical, everyday families. This therapeutic intervention offers families a valuable platform to comprehend one another and themselves better. It also helps them reflect on how their interactions and communication shape their relationships. As noted by one of the therapists from our context study, she said: \qt{whether an individual experiences a positive or negative family environment has varying effects on their emotional state. In the role of a family member, when someone is not in a `healthy' state, it often relates to their interactions within the family. Therefore, there are situations where it is crucial for family members to empathize and understand one another. When family members foster greater understanding of each other's feelings and needs, it invariably leads to positive outcomes}. 

\begin{figure*}[tb]
  \centering
  \includegraphics[width=8cm]{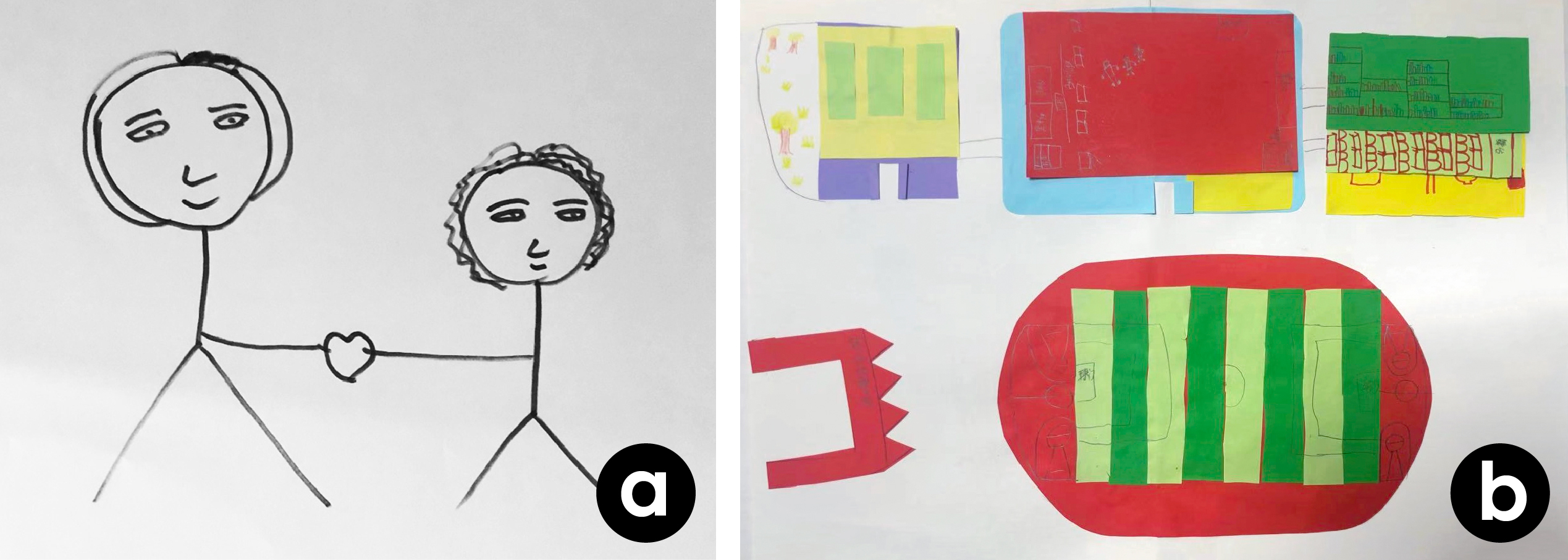}
  \vspace{-4mm}
  \caption{Family expressive arts therapy activities conducted by a therapist: (a) drawing; (b) collage.}
  \Description{Figure 2 contains two pictures. The first picture shows a mother making gestures with her daughter, showing her relationship with her mother; the second picture shows a child using collage to describe his school playground.}
  \label{fig:pilot}
\end{figure*}

As outlined by the therapists, family expressive arts therapy typically comprises the following phases: Preparation, Art Creation (including character-making, scene-making, and story-making), and Storytelling and Reflection (refer to Table~\ref{tab:stage}).
In their regular sessions, therapists draw from a diverse set of art resources and integrate therapeutic principles like projection, positive psychotherapy, play therapy, and narrative therapy.
A variety of traditional materials are employed to enhance the family's expression (for outcome examples, see \autoref{fig:pilot}).

\begin{table*}[tb]
\caption{Common Process of Family Expressive Arts Therapy.}
\centering
\vspace{-3mm}
\label{tab:stage}
\small
\begin{tabular}{m{3cm} m{6.5cm} m{4.5cm}}
\toprule 
Stage & Description & Therapeutic Meaning \\
\midrule
Preparation 
& The therapist initiates a dialogue with the client to understand their current feelings and needs, then collaborates to set session goals. 
& \begin{itemize}
\item Establishing trust and comfort; 
\end{itemize} 
\\
\begin{tabular}[l]{@{}l@{}}Art Creation \\(Character-making, \\Scene-making,\\Story-making)\end{tabular}
& Clients typically engage in developing characters, scenes and stories, using a variety of expressive materials, and the therapist usually observe them during the entire process. 
& 
\begin{itemize}
\item Facilitating nonverbal communication; 
\item Allowing for energy release;
\item Building a safe space;
\item Facilitates projecting one's feelings.
\end{itemize}
\\
Storytelling and Reflection 
& The therapist discusses the artworks with the clients, and allows the clients to tell stories through performances and narratives etc. Meanwhile, the therapist also ask various questions. 
& 
\begin{itemize}
\item Fostering connection between life and story;
\item Exploring their creativity;
\item Facilitating their self-reflection.
\end{itemize}
\\
 \bottomrule
\end{tabular}
\end{table*}

Upon being introduced to the image-based generative AI tools, the therapists envisaged their possible integration into the family expressive arts therapy. They believed that AI holds great potential if it can \textit{blend into, instead of replacing} physical materials. Namely, they considered that AI might lower the creative threshold for families. And the rich variations of AI-generated images might enrich families' nonverbal expressions of ideas and feelings. AI might be combined with different materials, such as using family-created artifacts as prompts for AI and using AI-generated images as elements for collage-making. These inputs from the therapists valuably clarified the current practice of family expressive arts therapy and were instrumental in shaping our study design.


%% file: texfiles/4-Methods.tex
\section{Methods}
Our research objective is to explore how parents and children engage in AI-infused multimaterial storymaking for expressive arts therapy (\textbf{RQ1}), how generative AI can be leveraged as expressive materials in this context (\textbf{RQ2}), and gather implications about how to support children, parents, and therapists in family expressive arts therapy (\textbf{RQ3}). Therefore, we conducted five weeks of AI-infused, multimaterial therapeutic storymaking sessions with seven families guided by a professional therapist, T1 (see \autoref{tab:expert}). These sessions were based on T1's existing practice as introduced above.

\subsection{Family Recruitment}
\label{sec:fff}
To recruit family groups, the therapist distributed digital recruitment flyers through her client channels. In addition, the research team also distributed the flyers through social networks. 
The recruitment flyer outlined our activity's aim as facilitating parent-child communication and strengthening their relationship. 
This aligns with the predominant needs expressed by the therapist's clients and the overarching objective of family-centered expressive arts therapies, as noted in~\cite{kerr2011family}. 
The flyer described the activity as ``collaboratively crafting stories with your children using a suite of engaging digital software'', without mentioning the incorporation of generative AI. Parents informed their children based on same information. 
Finally, we recruited 7 family groups, comprising a total of 18 participants~(8 were children aged 4-7, 7 were mothers, and 3 were fathers). Table~\ref{tab:freq} shows the specific participant demographics of the 18 participants from 7 family groups, all families were from the city of ShenZhen.
Our protocol received approval from the institutional research ethics board, and all names mentioned in this paper have been replaced with pseudonyms. 
Also, before participating in the activity, participants diligently read and acknowledged the informed consent form by appending their signatures. All families willingly took part in without compensation; the therapist was compensated with her regular hourly rate.

\begin{table*}[tb]
\caption{Demographics of Participant Families: Family Number indicates the specific family to which the participant belongs.}
\vspace{-3mm}
\label{tab:freq}
\small
\begin{tabular}{cccccc}
\toprule
ID & Family Number& Role & Gender& Age &Educational Level \\
\midrule
P1& F1& Mother& F & 38 & Bachelor\\
P2& F1& Daughter& F & 6 & Pre-school \\
P3& F2& Mother& F & 40 & Master \\\
P4& F2& Son& M & 7 & Primary \\
P5& F3& Mother& F & 40 & Bachelor \\
P6& F3& Father& M & 42 & Bachelor \\
P7& F3& Daughter& F & 7 & Primary \\
P8& F3& Son& M & 4 & Pre-school \\
P9& F4& Mother& F & 35& Master \\
P10& F4& Daughter& F & 6 & Pre-school \\
P11& F5& Mother& F & 34 & Master \\
P12& F5& Father& M & 35 & Master \\
P13& F5& Son& M & 5 & Pre-school \\
P14& F6& Mother& F & 45 & Master \\
P15& F6& Father& M & 48 & Doctor \\
P16& F6& Son& M & 6 & Pre-school \\
P17& F7& Mother& F & 39 & Doctor \\
P18& F7& Son& M & 7 & Primary \\

\bottomrule
\end{tabular}
\end{table*}

\subsection{Materials and Procedure}
\subsubsection{Materials}
A multitude of materials can afford diverse expressive properties to promote clients' expression through making~\cite{lazar2018making}. 
Thus, we introduced a diverse array of physical expressive materials, including colored pencils, colored markers, paint sticks, clay, non-woven fabric, yarn, drawing paper, and LEGO~\textsuperscript{\textregistered} classic~(see \autoref{fig:setup2}~\circled{c}).
Meanwhile, we encouraged children and their parents to bring children's favorite toys to our activities and incorporate these toys into their own stories, thereby imbuing their creative expressions with a personalized touch.

In addition to physical expressive materials, we introduced one of the most popular image-generative AI tools, Midjourney~\cite{midwebsite}, as a research probe due to several advantages: (1) customization and easy-to-use: it provides a broad range of customization features and a highly user-friendly interface~\cite{midad}. (2) speed and efficiency: it quickly generates diverse high-quality images without prolonged prompt setup~\cite{midad}.
In our study, we primarily utilized visual arts co-created by family members through the manipulation of physical materials as image prompts for generating AI images. Additionally, we enabled parents, children, and the therapist to employ their own verbal descriptions as text prompts. The incorporation of both verbal and non-verbal forms holds significant importance in the context of expressive arts therapy~\cite{malchiodi2003expressive}.

In order to quickly generate AI images during the activities, we prepared the Midjourney prompt guidelines based on our study in advance. We referred to~\cite{midguide} for guidelines on Midjourney prompt writing. For character and object generation, we use ``image URLs + subjects + descriptors of subjects + style descriptors + white background + parameters (aspect ratio: 3:2)''; for scene generation, we use ``image URLs + subjects + descriptors of subjects + style descriptors + parameters (aspect ratio: 16:9)''. Examples of the specific prompt guidelines can be found in APPENDIX.

\subsubsection{Procedure}
\begin{figure*}[tb]
  \centering
  \includegraphics[width=8cm]{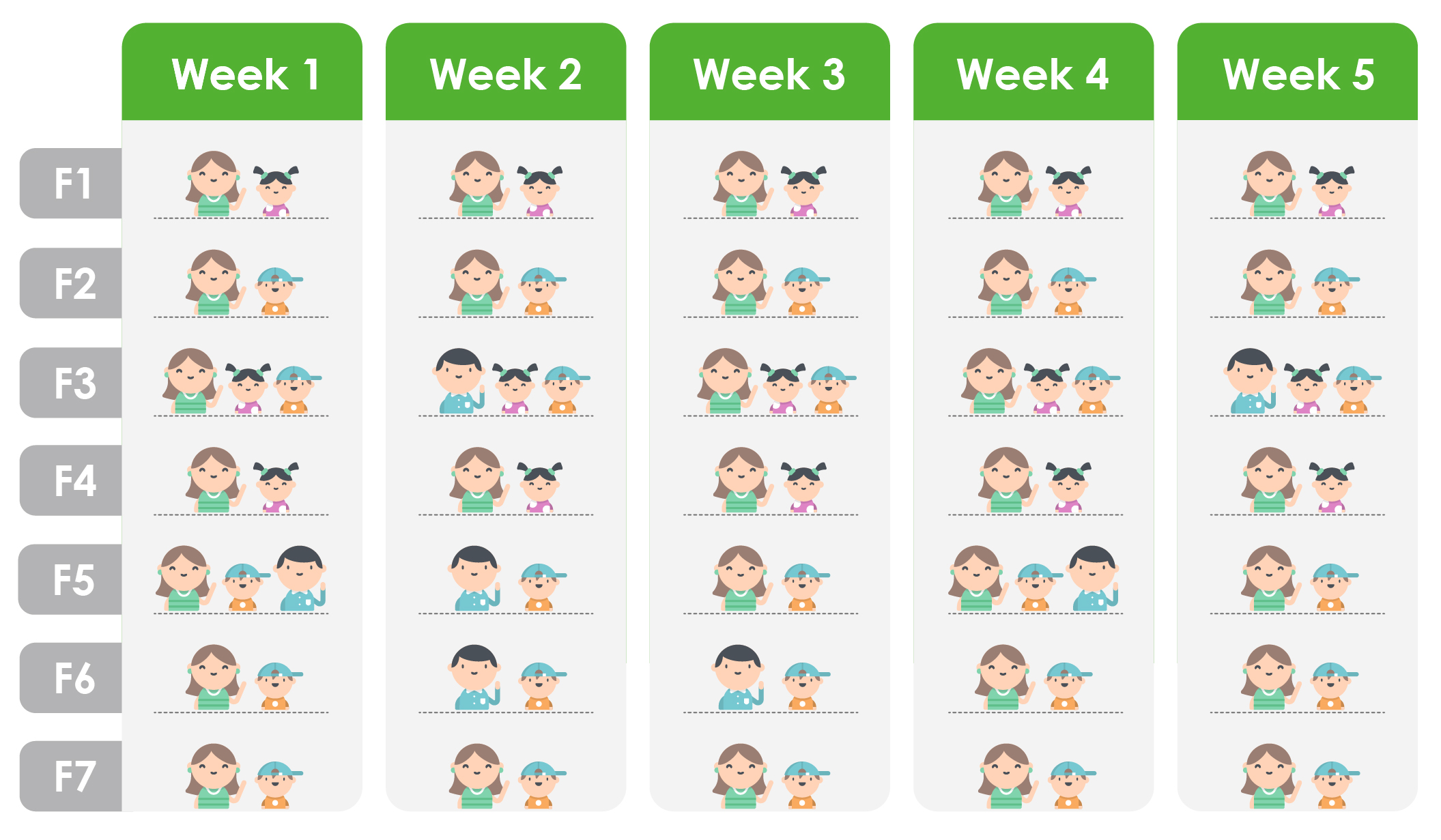}
  \vspace{-4mm}
  \caption{Participation status of family members in 7 groups of families}
  \Description{Figure 4 shows participation status of family members in 7 groups of families.The activity comprises seven distinct family groups, each participating in a series of five-week sessions. Family 1 consists of a mother and daughter, Family 2 involves the current son, Family 3 includes both parents along with a son and daughter, Family 4 comprises a mother and daughter, Family 5 consists of a mother, father, and son, Family 6 comprises a mother, father, and son, and Family 7 involves a mother and son.}
  \label{fig:setup1}
\end{figure*}

Between April and July 2023, each family participated in five storymaking sessions guided by T1. Before the activities, the therapist conversed with families to gain insights into the needs and goals of families (serving as the Preparation phase in \autoref{tab:stage}). These sessions, each an hour long, were spaced at least a week apart. This grants the time necessary for the therapeutic intervention to unfold and also enables us to gather participants' insights over an extended period. While the significance of long-term therapeutic approach is underscored in research~\cite{greenwood2007process}, it is infrequently accommodated in HCI studies. \autoref{fig:setup1}~\circled{a} shows the participation of each family in the five sessions. 

\begin{figure*}[tb]
  \centering
  \includegraphics[width=\linewidth]{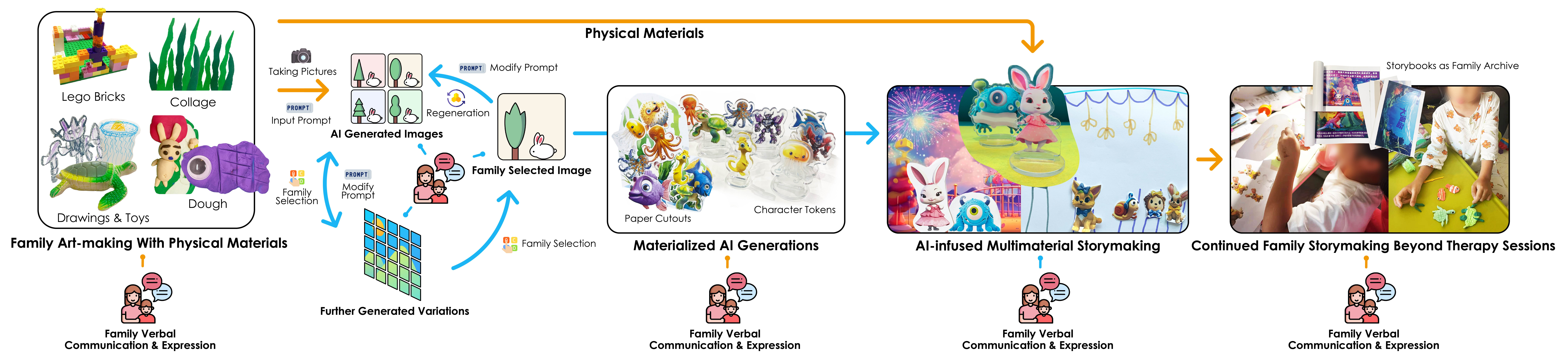}
  \vspace{-4mm}
  \caption{Illustration of how families combined analog materials and generative AI in their creative process.}
  \Description{This figure shows the AI-infused multimaterial storymaking study process. From the left we can see parents and children took the lead in the creative process through physical art-making, such as Lego or their own toys, and verbal communication. The researcher helped the family document the artifacts they co-created into photos, and used them as input images for generative AI. The family's verbal articulation about their artifacts was used as text prompts. Four images were generated each time, and the family could choose the one they preferred, or provide the researcher with feedback for content or style adjustments to create more variations. They could also modify their artifact or iterate their verbal articulation as updated AI inputs for regeneration. Such (re-)generation cycles could run many times to grant families flexibility and autonomy in exploring abundant variations until they identify the most suitable visual expression in terms of both content and art style.}
  \label{fig:studydesign}
\end{figure*}

During each session, in addition to the therapist and the family, a researcher was providing technical support, including operating the generative AI.
Primarily, parents and children took the lead in the creative process through physical art-making and verbal communication, as~\autoref{fig:studydesign} shows. The researcher helped the family document the artifacts they co-created into photos, and used them as input images for generative AI. The family's verbal articulation about their artifacts was used as text prompts.
Four images were generated each time, and the family could choose the one they preferred, or provide the researcher with feedback for content or style adjustments to create more variations.
A tablet was provided to the family to view and select generated images on the spot.
They could also modify their artifact or iterate their verbal articulation as updated AI inputs for regeneration.
Such (re-)generation cycles could run many times to grant families flexibility and autonomy in exploring abundant variations until they identify the most suitable visual expression in terms of both content and art style.

AI-generated images were employed as intermediate expressive materials, instead of final products, as shown in~\autoref{fig:studydesign}. 
With the researcher's help, these AI images were \textit{materialized} into printed scenes, character paper cutouts, and physical tokens for further co-creation. 
Families could then combine these materialized AI generations with other analog materials for further AI-infused multimaterial storymaking both within and beyond the therapy sessions.

The therapist's role concerns facilitating family co-creation and encouraging their verbal communication and expression.
An activity framework was planned based on the therapist's current practice to aid parents and children in their developmental journey~(see \autoref{fig:setup2}~\circled{a}). These activities, as described below, were distributed in varying proportions throughout the five sessions and were not conducted in a strict sequential order:

\textbf{Character Making.}
In this activity, the expressive arts therapist could guide parents and children to try different physical expressive materials or use own toys to create story characters that might help to project their thoughts and feelings through personal metaphor~\cite{perrow2008healing}. Once they were created, the researcher employed a mobile phone to take photos, which were subsequently uploaded to Midjourney following image pre-processing. Subsequently, AI-generated images were created based on the verbal descriptions provided by the family members or the therapist. Then, parents and children could choose desirable images from four variations. Finally, the therapist also encouraged family members to name the characters, which was a strong assertion of one’s presence and agency~\cite{david2020models}.

\textbf{Scene/Object Making.} In this activity, children and parents could make stories' objects and scenes by drawing, building Lego, kneading clay, and collaging according to their own interests. The therapist guided family members to infuse their personal life experiences into the fabric of the story's scenes and objects, thereby offering insight into various aspects of one's life and outlook.

\textbf{Story Making and Storytelling.} In this activity, parents and children co-created story plots. 
According to a construction model for therapeutic story~\cite{perrow2008healing} following a three-part framework of `metaphor', `journey' and `resolution', 
family members facilitated by the therapist created their own story plots through collages and narratives. This activity maintained a relatively open structure, leaving abundant space for creativity, and thus children and parents are encouraged to take a proactive role in this activity.

\textbf{Story Sharing and Reflection.} 
In this activity, parents and children shared their own storybooks created by the generative AI tool, and the therapist could engage in parent-child discussions by asking questions to facilitate their reflection. 

\subsubsection{Backstage Preparation}
\begin{figure*}[tb]
  \centering
  \includegraphics[width=\linewidth]{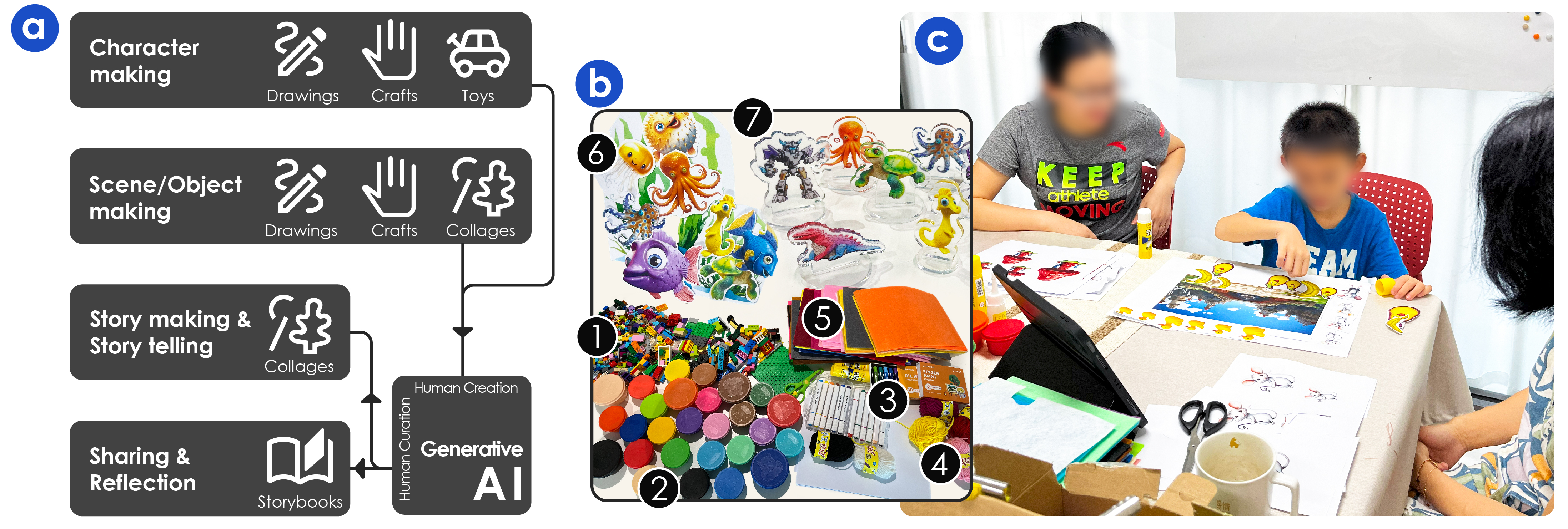}
  \vspace{-4mm}
  \caption{Study setup: (a) overview of the four activities; (b) diverse expressive materials (1-5: LEGO, colored clay, painting mediums, yarn, non-woven fabric; 6: paper cutouts of AI; 7: character tokens); (c) activity participants(mother: left, children and therapist: right).}
  \Description{Figure 5 shows three pictures. The leftmost picture shows an overview of the four activities, including character making, scene/object making, storytelling/storymaking, sharing and reflection; the middle picture shows diversity physical materials and AI generated materials. These include 1-5: LEGO, colored clay, painting mediums, yarn, non-woven fabric; 6: paper cutouts of AI; 7: character tokens; the rightmost picture shows activity participants (mother: left, children and therapist: right ), mother and child are telling stories using AI-generated paper cutouts.}
  \label{fig:setup2}
\end{figure*}

Throughout the entire process, our researchers undertook backstage preparations for the sessions.
First, retouching tools were employed to perform tasks such as cutting out, mirroring, and refining characters and objects. These characters and objects were then arranged in diverse sizes on A4 paper and subsequently printed~(see \autoref{fig:setup2}~\circled{b}~(6)). 
Also, AI-generated scene images were printed after a session. These paper cutouts of AI generated content were employed as expressive materials in the next session. 
In order to probe the continued, real-life impact of storymaking artifacts outside the therapeutic sessions, we offered family-created, AI-generated character tokens for children to take home after the sessions~(see \autoref{fig:setup2}~\circled{b}~(7)). 
Finally, following the handmade multimaterial storybook frames and finalized storylines co-created by the families in the sessions, our researchers and the therapist used retouching tools to organize and arrange digital storybooks. These were later printed out for each family. The examples of the storybooks are detailed in APPENDIX.

\subsubsection{Data Collection and Data Analysis}
All sessions were audio- and video-recorded.
At the end of each session, we conducted a 5-minute interview, mainly to collect the experiences of parents and children about the session. 
After the third session, an intermediate interview planned for thirty minutes was conducted with each family to collect their feedback and experiences halfway through the study. 
Upon concluding all the sessions, we held a one-hour final interview with each family, in which the outcome artifacts from all five sessions, including analog artworks and AI-generated images, were used as stimuli.
We also encouraged the families to send us images and comments via instant messaging applications to share their additional thoughts or how they engaged with the storymaking artifacts outside the sessions. 
At the end of the intermediate and final interviews, we invited the therapist to join the discussion to share her observations and suggestions.

Our analytical focus was on the interactions between families and AI-infused multimaterials and their therapeutic meanings of AI. All the interview data were transcribed verbatim. 
Following a thematic analysis approach~\cite{braun2006using}, 
two researchers conducted collaborative inductive coding. They initially annotated the transcript to identify relevant quotes, key concepts, and preliminary patterns in the data. These findings were further developed through regular discussions, leading to a detailed coding scheme aligned with the research objectives. Quotes were then coded and clustered into a hierarchy of emerging themes, continually reviewed, and refined in recurrent meetings, where exemplar quotes were also selected for presenting each theme and sub-theme. 

Alongside this, the team reviewed and annotated the session videos, keeping both the research questions and the thematic analysis in perspective. We collected the video segments that could act as evidence or exemplars for the thematic analysis results, especially those highlighting behaviors of interest among children, parents, and therapists. In addition, we supplemented our analysis with photographic documentation of family-created storymaking artifacts from the sessions, and utilized relevant expressive arts therapy theories to deepen our insights into the data.

To further enrich our data interpretation with broader therapeutic insights, we conducted a two-hour expert review session with four therapists (T1-T4; T1 directed all storymaking sessions) whose demographics and expertise are detailed in~\autoref{tab:expert}. We provided a comprehensive presentation with images of both the artifacts and the process of the activities. Representative video clips from our analysis were also showcased. Therapists were prompted to pose questions and offer commentary to both researchers and to each other. The discourse was primarily anchored around our research questions.
In line with our earlier approach, we employed the thematic analysis method, as referenced in ~\cite{braun2006using}, to dissect the therapists' commentaries.

\begin{table*}[tb]
\caption{Demographics of Participant Therapists: Experience refers to the number of years engaged in expressive arts therapy; The Number of Case refers to cases related to family therapy}
\label{tab:expert}
\vspace{-3mm}
\small
\begin{tabular}{ccccccc}
\toprule
ID & Age & Gender & Experience & Education Level& Major & The Number of Case\\
\midrule
T1& 48& F& 5 & Master & Expressive Arts Therapy &57\\
T2& 32& F& 7 & Master & Art Psychotherapy &100+ \\
T3& 39& F& 9 & PhD & Art Therapy &100+ \\
T4& 33& F& 4 & Master & Expressive Arts Therapy &50+ \\
\bottomrule
\end{tabular}
\end{table*}

%% file: texfiles/Findings.tex
\section{Findings}
\subsection{Outcomes and Process of Family Engagement in AI-infused Multimaterial Storymaking (\textbf{RQ1})}
In this section, we offer an overview of how the children and parents engaged in storymaking-based expressive arts therapy combining generative AI and analog materials \textbf{RQ1}. This overview helps us contextualize the creative process of the families and the therapeutic meaning behind it. As illustrated in \autoref{fig:finding1}, their journey has been structured in four major activities:

\begin{figure*}[tb]
  \centering
  \includegraphics[width=\linewidth]{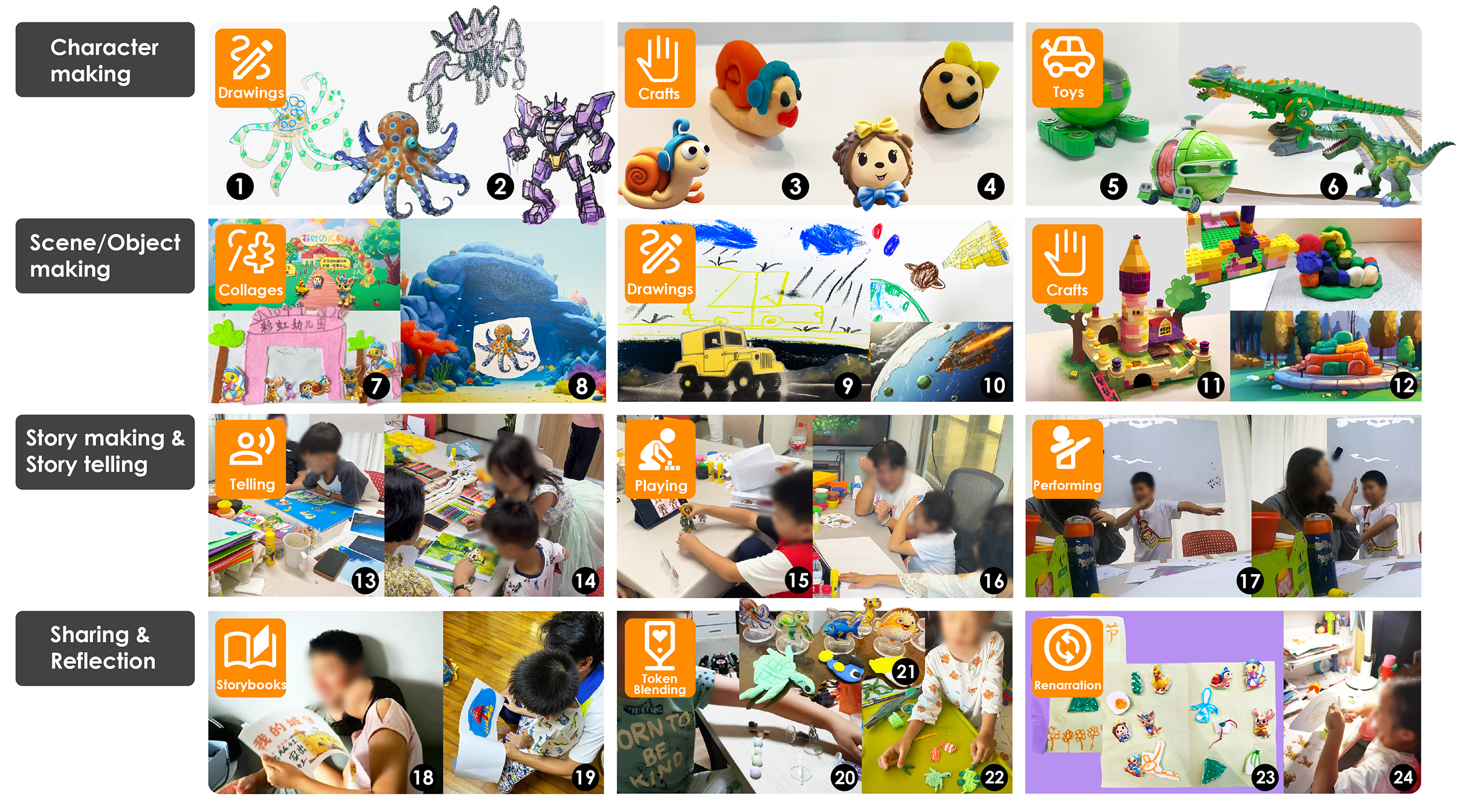}
  \vspace{-4mm}
  \caption{Outcomes and process of family engagement in AI-infused multimaterial storymaking in four major activities.}
  \Description{Figure 1 portrays a family employing a palette of vibrant colors, specifically blue and green, to craft an eight-legged blue-ringed octopus on paper. Simultaneously, an AI-generated cartoon image emerges, featuring a one-eyed rendition of the captivating creature. In Figure 2, a child's artistic expression unfolds as spontaneous brushstrokes and formalized lines converge to give life to a regal purple Transformers. AI generated a cartoon purple Transformers, harmonizing with the child's imaginative strokes. In Figure 3, a child explores the realm of colorful structures, utilizing playful hues such as red, blue, yellow, and black. The juxtaposition of a snail adorned with blue headphones, alongside its AI-generated counterpart with oversized eyes and matching headphones, showcases the embodiment of a childlike aesthetic. Figure 4 captures the child's uninhibited use of differently colored clay(a hedgehog bedecked with a yellow rope knot.). Figure 5 features a green watermelon-shaped toy car, mirrored by an AI-generated image that seamlessly complements the child's creation. Figure 6 presents a menacing green dinosaur, mouth agape, brought to life through AI wizardry. Figure 7 unfolds a delightful scene where children, through collage, express the joyous tapestry of their kindergarten life. The children used soft pastel-colored non-woven fabric to make the scene of the kindergarten which placed two symmetrical trees at the entrance of the kindergarten. Different AI-generated characters of paper cutout are neatly arranged together. In Figure 8, a child utilized collage to depict a blue-ringed octopus dwelling in the profound, shadowy depths of an oceanic cave surrounded by vibrant seaweed. Figure 9 employs dark tones and scenes to evoke a profound emotional resonance, depicting blue clouds, pouring rain, yellow trucks, and black roads—a visual testament to heavy, depressing, or gloomy feelings. Figure 10 showcases a child's artistry with an eclectic blend of colors, fashioning a symmetrical and balanced Lego castle and clay park. Figure 13 shows a child telling a story to his parents. He uses artificial intelligence paper cutouts and blue cardboard to immersively tell stories that happen under the sea. Figure 14 shows two siblings sharing their own stories with each other. Figure 15 shows the child happily playing with the Transformers he brought from home and the character tokens generated by AI and improvising the story. Figure 16 presents a role-playing activity which a child, aided by AI-generated character cutouts, engages in role-playing with their mother, placing the character cutout on their forehead. Figure 16 shows that the daughter and mother were playing roles. The daughter was immersed in telling what happened in the story by placing the AI-generated paper cutout on her forehead. Figure 17 shows children using body performance to tell stories about Godzilla and Transformers with their parents. Figures 18 and 19 depict heartwarming moments through sharing physical storybooks with their father in their home. In Figure 20, a child is depicted integrating AI-generated character tokens with personal toys at home, facilitating interactions between the AI-generated characters and their toys. Figures 21 and 22 illustrate the imaginative prowess of children as they craft analogous characters from colored clay inspired by AI-generated character tokens. Moving forward to Figures 23 and 24, these snapshots capture the creative journey of children meticulously crafting their own storybooks using a blend of artificial intelligence materials and traditional art supplies. This process serves as a meaningful connection between their past and future energies through the medium of artistic creation.}
  \label{fig:finding1}
\end{figure*}

\textbf{Character Making.} Most children and parents used drawing or clay-kneading combined with the generative AI to create their characters~(see \autoref{fig:finding1}~\circled{1}, \circled{2}, \circled{3} and \circled{4}). Some children also incorporated their toys as story characters, echoing the idea of externalized metaphors~\cite{perrow2008healing}. For example, P8 used a dinosaur toy as an obstacle metaphor~(see \autoref{fig:finding1}~\circled{6}). At the same time, certain story characters could project children's emotional and inner thoughts: \qt{[...] a consistent character [...] could be regarded as a projection of the child's own self […] For the child, the primary focus is on how they project these characters into their own stories~(T3)}. 

\textbf{Scene or Object Making.} Besides drawing~(see \autoref{fig:finding1}~\circled{9} and \circled{10}), LEGO~\textsuperscript{\textregistered}-building~(see \autoref{fig:finding1}~\circled{11}), and clay-kneading~(see \autoref{fig:finding1}~\circled{12}), children made their own scenes by mixing the paper cutouts of AI-images with physical materials~(wool, colored clay or non-woven fabric). For instance, P10 mixed the AI-image cutouts with non-woven fabric to make collages, which emerged as a common way for scene-making across families~( \autoref{fig:finding1}~\circled{7}). In constructing scenes/objects, most children added their experienced real-life scenarios, such as hospitals or kinder gardens into the image-prompts or artifacts. As noted by related theory~\cite{hinz2019expressive}, their arrangement and structuring of scenes could reflect their emotional experiences from those scenarios.

\textbf{Storymaking and Storytelling.} As an emerging creative example, P13 created stories using a blue cardboard to represent the underwater world, on which he moved the character cutouts of both AI images and drawings to act out his stories on-the-spot~(see \autoref{fig:finding1}~\circled{13}).
Role-play was frequently used by the therapist to promote storymaking. For example, T1 prompted P1 and P2 to exchange their character paper cutouts, allowing them to converse and develop the story while swapping their roles~(see \autoref{fig:finding1}~\circled{16}), offering potential benefits of perspective taking~\cite{jennings2005embodiment,ahmadpour2023understanding}. P18 created stories by playing with character tokens and personal toys~(see \autoref{fig:finding1}~\circled{15}), which could enable children to communicate to self and make meaning of their experiences~\cite{schaefer2011foundations}. Finally, bodily performance was founded in our sessions~(see \autoref{fig:finding1}~\circled{17}), which has been deemed to facilitate embodied exploration, rehearsal, and demonstration of story ideas or creative slotuions~\cite{jones1991dramatherapy}.

\textbf{Sharing and Reflection.} Besides reading and discussing their storybooks during the sessions, children also shared, developed, and re-narrated their stories outside or after the sessions. Namely, many children reviewed their storybooks with family members at home~(see \autoref{fig:finding1}~\circled{18} and \circled{19}); P13 proudly shared his storybook and making process in a kindergarten weekly sharing session~(see \autoref{fig:teaser}~\circled{d}). As P11 added: \qt{[...] When P13 brought up the topic of AI storybooks with fellow children, they didn't know what he was talking about. [...] But he told his friends with pride `I am going to AI today!'} Moreover, children improvised new stories by playing with character tokens and their toys at home and made new crafts using the tokens~(see \autoref{fig:finding1}~\circled{20}, \circled{21} and \circled{22}). Children used AI-generated materials to continue crafting new storybooks at home~(see \autoref{fig:finding1}~\circled{23} and \circled{24}). P10 used both paper cutouts and materials from life: \qt{After the Dragon Boat Festival [...] I took out the paper cutouts from the drawer and encouraged her to make a storybook about it [...] I saw her cutting these papers. Then, I told her that there was foam in the express package, and she said that this foam could be used to make characters pop up~(P9)}.

\subsection{Examining AI-infused Material Interactions through the Lens of ETC (\textbf{RQ2})}
\begin{figure*}[tb]
  \centering
  \includegraphics[width=\linewidth]{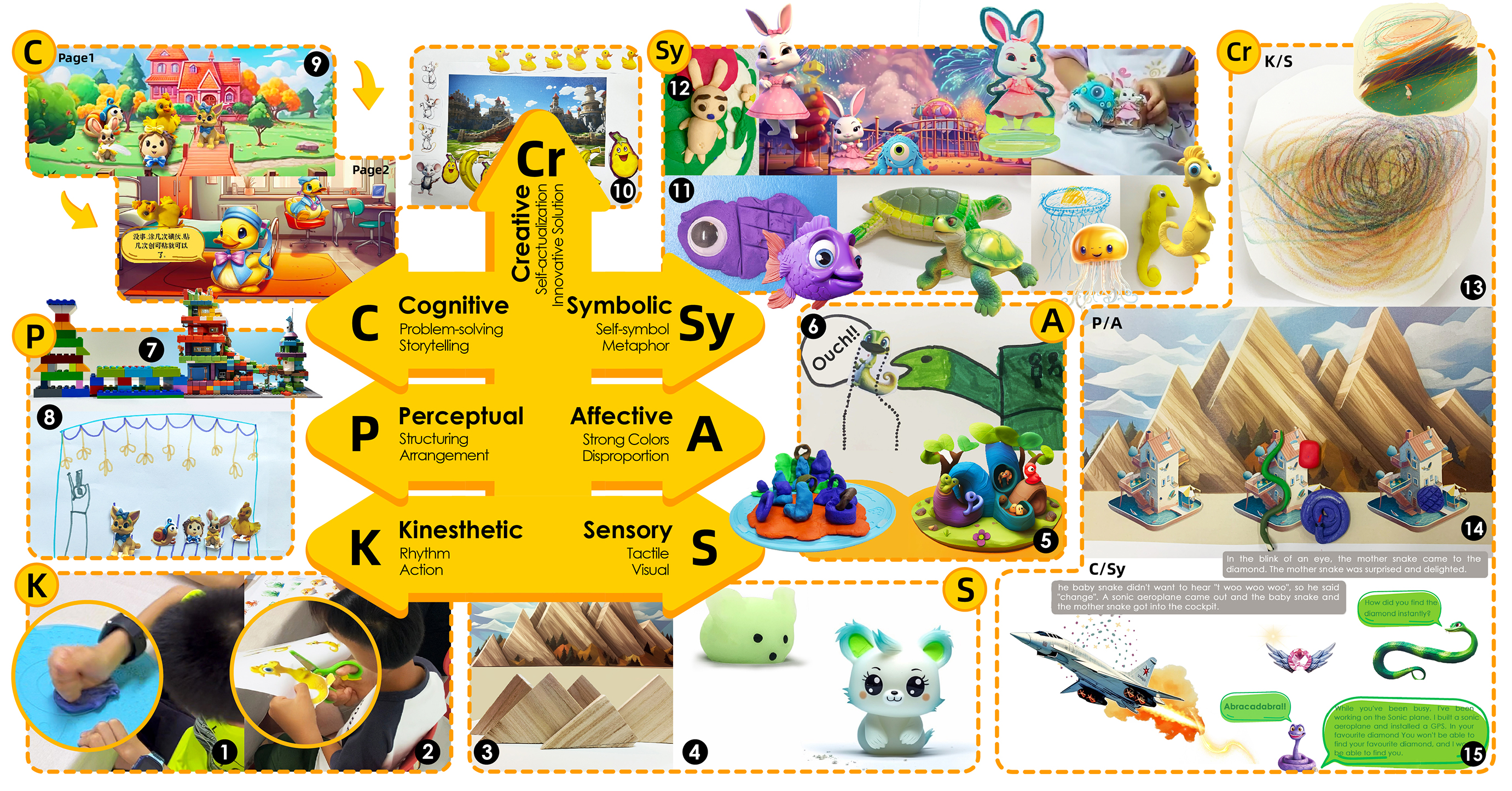}
  \vspace{-4mm}
  \caption{AI-infused material interactions from the lens of Expressive Therapies Continuum (ETC).}
  \Description{Figure 1 shows the scene of children hitting the clay heavily with their hands to make a sound, thereby achieving a cathartic effect. Figure 2 shows a child using scissors cutting out AI printed yellow seahorse character. In Figure 3, a child was consistently arranging boards with wood-textured surfaces, revealing the occlusion relationship between the boards. The AI then generates corresponding images featuring mountain peaks, drawing inspiration from the wood-textured planks. In Figure 4, a soft, translucent toy is depicted. The AI has generated a crystal-clear cartoon hamster character, drawing inspiration directly from this particular toy. Figure 5 shows a combination of different vibrant colors, rule-breaking shapes and form distortions, which evoked children' emotion. Figure 6 shows a huge green snake alongside a small snake (AI cutout) with exaggerated tears to depict fear. Figure 7 shows a child organizing the LEGO® with a high degree of form potential in a visual order (own garage and house).In Figure 8, a child arranged a row of AI characters within a classroom setting adorned with an abundance of yellow flowers. Figure 9 shows a scene that took place in a kindergarten. A group of AI-generated characters were playing outdoors. Suddenly a small animal was injured. The child employed problem-solving skills to address various injuries that arise within her storybook. Figure 10 shows the scene that took place at the racing track. child employed AI-printed collages to distribute banana to different characters logically through selection, organization and calculation.Figure 11 shows the child creating a purple fish with big eyes, his own green toy and AI-generated cartoon images, an omelette jellyfish and a yellow seahorse. They use these characters as self-symbols.Figure 12 shows the child using handwork and artificial intelligence to create a little rabbit character lying on a bed full of flowers. AI helped generate a little rabbit wearing a pink skirt and a pink bow. This little rabbit was featured in her story. Used as a self-symbol. At the same time, this character is integrated into her real life in the form of a symbolic physical token. Figure 13 shows a child held multiple colored pens and let body movements guide the creation of colored traces on the paper without specific intent, no specific shape. The child then used sensory perception to liken them to tornadoes. Figure 14 shows the child placed cutouts of cartoon-style houses and mountains with diving platforms in a pleasing order atop a printed background. The child then creatively turned this paper collage into a ``stage'' to hold the clay characters (a green stripe little snake and a little purple snake) celebrating their family gathering with joy and contentment. Figure 15 shows the baby purple snake crafted and piloted a magical plane on a quest to help its green snake mother find a missing diamond with wings.}
  \label{fig:etc}
\end{figure*}

In this section, we concretely examine how generative AI—when seen as expressive materials used along with other analog materials—could support the material interactions along different dimensions of the Expressive Therapies Continuum (\textbf{RQ2}). 
Material interactions play a pivotal role in the therapeutic outcomes and the overall experience of the clients~\cite{lazar2018making}. 
As a foundational theoretical framework, Expressive Therapies Continuum~(ETC) describes four dimensions along which clients interact with materials to process information and form images from simple to complex~\cite{hinz2019expressive}.
We employed ETC as a theoretical lens to examine AI-infused material interactions concretized by our data, to inform future usage of generative AI for expressive arts therapy~(see \autoref{fig:etc}).

\textbf{Kinesthetic/Sensory Level.} At the Kinesthetic/Sensory Level, the processing of information commences at a preverbal stage~\cite{hinz2019expressive}. 
The Kinesthetic component encompasses the sensations that provide individuals with information about bodily rhythms, movements, and actions~\cite{hinz2019expressive}. 
The Sensory component encompasses information derived from visual, auditory, and tactile channels~\cite{hinz2019expressive}. 
On the Kinesthetic component, we found that children engaged in physical manipulation and handling of various materials through movement and actions, such as pounding clay with their hands~(P4, see \autoref{fig:etc}~\circled{1}), cutting out printed AI-generated objects/characters~(P13, see \autoref{fig:etc}~\circled{2}), moving rhythmically with acrylic standees of AI characters~(P18), pushing~(P2, P4, P16), stabbing and cutting clay~(P4), and cutting non-woven fabric~(P2, P10, P13). 
According to related theories, these activities can facilitate individuals' release of energy, physical arousal, and tension~\cite{luria2020destruction}. It also promotes relaxation for clients~\cite{slater2004healing}. As noted by P5, the move of shaping the clay for character making (as inputs for AI) was \qt{quite relaxing}.
On the Sensory component, we found that families engaged the sensory qualities of various materials through tactile, visual, and auditory elements. And they intended to bring some of the appreciated visual or textural qualities into the AI images.
For example, P4 input a board with wooden texture~(see \autoref{fig:etc}~\circled{3}), and also P17 input what she described as a soft, translucent texture of a toy~(see \autoref{fig:etc}~\circled{4}). According to related theories~\cite{lusebrink1990imagery}, these activities can enhance broader attention to the present moment, encourage self-soothing, and express inner sensation.

\textbf{Perceptual/Affective Level.}
The Perceptual/Affective Level concerns information processing and image formation either verbally or nonberbally~\cite{hinz2019expressive}. 
The Perceptual component features the regular arrangement, ordering, or completion of form elements~\cite{hinz2019expressive}. 
On the Affective component, information processing and image formation can be mediated by emotional or raw expressions using unbalanced proportions or vibrant colors~\cite{kagin1978expressive}.
On the Perceptual component, we found that family members tended to create images through ordered arrangement, and regular patterns using structural materials~(e.g, wood, LEGO~\textsuperscript{\textregistered}).  
For instance, P16 organized the LEGO~\textsuperscript{\textregistered} bricks into a clear structure to create a building~(see \autoref{fig:etc}~\circled{7}). 
Children also created regular formations of multiple characters on their collages~(P4, P10, and P16), e.g., P10 formed a row of AI characters in the classroom scene~(see \autoref{fig:etc}~\circled{8}). 
Such perceptual aspects in expressive making could increase structural awareness and foster a satisfying internal state~\cite{ulman2001art,hinz2019expressive}. 
On the side of the Affective component, we found that some children picked vibrant or strong colors in their clay-kneading or collages~(P4, P8, P16 and P18). P8 mobilized different vibrant colors and rule-breaking shapes in his clay artifact to reflect his joyful state~(see \autoref{fig:etc}~\circled{5}). P7 drew a huge snake alongside a small snake (AI cutout) with exaggerated tears to depict fear~(see \autoref{fig:etc}~\circled{6}). Such affective aspects bring therapeutic benefits by allowing families to explore and access emotions~\cite{hinz2019expressive}.

\textbf{Cognitive/Symbolic level.}
The Cognitive/Symbolic level is the most complex and sophisticated, involving planning, cognitive action, and intuitive recognition~\cite{hinz2019expressive}. 
The Cognitive component involves deliberate planning, logical/realistic thinking, problem-solving, and storytelling~\cite{lusebrink2015expressive}.
The Symbolic component involves the intuitive, mythical, and self-relevant concept creating, and the articulate expression and resolution of symbolic elements~\cite{kagin1978expressive}.
On the Cognitive component, we found that children could plan and sequence their story plots with their parents through narratives and collages~(P2, P7, P10, P13, P16, and P18). 
F16 employed AI-generated paper cutouts to select, organize, calculate, and distribute food to different characters logically through the medium of collage~(see \autoref{fig:etc}~\circled{10}). P10 utilized problem-solving skills to address the injury of a character in her story~(see \autoref{fig:etc}~\circled{9}). These cognitive aspects in storymaking can benefit planning, cause-and-effect thinking, and problem-solving skills~\cite{hinz2019expressive}.
On the Symbolic component, we found that all children created self-relevant characters to connect stories and individual experiences through metaphoric symbols. For instance, P2 created a little rabbit character using handcrafting and AI, which was used as a self-symbol in her stories. Meanwhile, this character was incorporated into her actual life in the form of a symbolic physical token~(see \autoref{fig:etc}~\circled{12}). Each member of Family F5 had their own character as a self-symbol, and the tortoise toy seen as a friend by P13 was also added into the story through AI~(see \autoref{fig:etc}~\circled{11}). P11 mentioned that the most impressive experience was that the child made a \qt{purple fish} out of colored clay: \qt{P13 said at the time that he turned into a fish}. P11 added, \qt{he said that the seahorse is the mother and the jellyfish is the father}. Such symbolic aspects help children realize personal meaning and enable parents to understand their children through personal symbols~\cite{kagin1978expressive}.

\textbf{Creative level.}
The Creative level involves innovative and resourceful interactions with the environment and expressive materials, leading to creative expression and self-actualizing experiences~\cite{hinz2019expressive}. 
It can ``bring forth something new''~\cite{hinz2019expressive} within the expressive process across all levels of ETC. 
Using P4 and his family as an example, at the Kinesthetic/Sensory Level,
P4 held multiple colored pens and let body movements guide the creation of colored traces on the paper without specific intent. P4 then used sensory perception to liken them to tornadoes~(see \autoref{fig:etc}~\circled{13}); 
at the Perceptual/Affective level, P4 placed cutouts of houses in a pleasing order atop a printed background. He then creatively turned this paper collage into a ``stage'' to hold the clay characters celebrating their family gathering with joy and contentment~(see \autoref{fig:etc}~\circled{14}); 
at the Cognitive/Symbolic level, 
during the evolution of the story, P4 crafted and piloted a magical plane on a quest to help his mother find a missing diamond (see \autoref{fig:etc}~\circled{15}), unveiling both the inventive and valiant facets within P4 as noted by the therapist.
Such creative moments ignited along the process could sparkle the experience of innovating and self-actualizing~\cite{hinz2019expressive}.

\subsection{Surfacing the Therapeutic Meanings of Generative AI as Expressive Materials (\textbf{RQ2})}
In this section, we aim to explicate the therapeutic meanings of generative AI as expressive materials in family expressive arts therapy. 
Our findings revealed three themes~(see Table~\ref{tab:finding2ai}): (1) AI as empowerment for families; (2) AI as connection between physical and digital, life and stories; (3) AI as co-creator. 

\subsubsection{\textbf{AI as Empowerment for Families}}
We found that employing AI as expressive materials holds two forms of empowerment for family expressive arts therapy:

\textbf{Lowering the Creative Threshold and Increasing Creation Efficiency.} 
We found that AI could lower the creative threshold by enhancing the aesthetic quality of creation. For instance, P1 elucidated the appeal of AI-generated outcomes that exceeded expectations, especially for children: \qt{P2 said that the tool was so magical, which made the rabbit so beautiful}. This helped children boost expressive confidence and focus on creative expression rather than art-making skills: \qt{children may become more confident [...] knowing that the quality of their drawing didn't matter since AI can enhance images~(P17)}. T1 also explained why AI can enhance expressive confidence: \qt{they might be less concerned about the aesthetic perfection of their creation. This could diminish the feeling of intimidation and heighten their engagement}. Furthermore, AI increased families' creation efficiency. For example, as P3 appreciates, \qt{if I were to create an oil painting like this, it would demand a lot of time and effort. Yet, exceptional results can be rapidly delivered just by using two wooden blocks}. 
Increased efficiency could facilitate clients' improvisation and creative tendencies~\cite{barclay1999magic}.


\textbf{Unconditionally Accepting Families' Imagination and Affording Endless Variations.} 
First, AI could unconditionally accept the imaginative concepts of families as prompts. For example, P11 noted that: \qt{he might envision such a scene in his mind but struggle to articulate it due to his current language capability. Nevertheless, these images support him to imagine}.
As P14 experienced, \qt{I can't turn the things in my head into a specific look that I want, but AI can help.}
Similarly, to P1, \qt{the images here feel directly generated from what I imagined [...] which is quite magical.}
Such unconditional embracing of families' imaginations could foster their creative self-expression, and psychological self-healing and renewal~\cite{mcniff1992art}.

Second, AI can provide endless variations and hence offer families abundant expressive possibilities: \qt{AI presents many options but they all fall under a specific category~(P3)}. When facing the varying options generated by AI, some children: \qt{engaged in discussions with their parents to make joint choices [...] This revealed the interaction between parents and children during this selection process~(T1)}. In addition to rich options, P17 mentioned that AI might create surprises when merged with traditional materials: \qt{manipulating colored clay or constructing with Lego can be even more unpredictable, yielding unexpected surprises. What you end up with might not exactly match your initial concepts, that's perfectly acceptable}. 
The rich creative possibilities are deemed as empowerment to exploration~\cite{brandle2023empowerment}. 

\begin{table*}[tb]
\caption{Summary of the Therapeutic Meanings of Generative AI as Expressive Materials.}
\centering
\vspace{-3mm}
\label{tab:finding2ai}
\small
\begin{tabular}{m{3.5cm} m{3cm} m{7.5cm}}
\hline
Themes & Description & Example Quotes \\
\hline
\multirow{4}{*}{\begin{tabular}[l]{@{}l@{}}AI as Empowerment \\ for Families\end{tabular}}
&
Lowering the creative threshold and increasing creation efficiency;
&
\qt{children may become more confident [...] knowing that the quality of their drawing didn't matter since AI can enhance images~(P17)}; 
\\
&
Unconditionally accepting families' imagination and affording endless variations.
&
\qt{I can't turn the things in my head into a specific look that I want, but AI can help~(P14)}.
\\
\hline
\multirow{2}{*}{\begin{tabular}[l]{@{}l@{}}AI as the Connection \\between Physical and Digital,\\Life and Stories\end{tabular}}
&
Fusing and harmonizing analog and digital artifacts; 
&
\qt{once I created an artifact, and P4 crafted another, it allowed us to combine both creations into a unified image~(P3)};
\\
&
Generating symbols that can permeate life and stories.
&
\qt{she has immersed herself in this role because [...] she relates her current state to the character of the little rabbit~(P1)};
\\
\hline
\multirow{5}{*}{AI as Co-creator}
&
Family-AI mutual interpretations;
&
\qt{what fascinates me is that he not only selected the AI-image but
also interpreted its unexpected part as jeweled bananas~(T4)};
\\
&
Family-AI comprise and authorship.
&
\qt{initially, we probably had a preconceived notion, but when AI-generated images deviated from our expectations, we found this variation to be acceptable~(P17)};

\qt{AI provided several images and allowed them to choose one that closely mirrored their own creation. This may be considered as respecting their ideas~(P5)}.
\\
\hline
\end{tabular}
\end{table*}

\subsubsection{\textbf{AI as Connection between Physical and Digital, Life and Stories.}} 
In this study, we found that AI as expressive materials can foster the connection between the physical and digital world, as well as real life and stories, in twofold meanings:

\textbf{Fusing and Harmonizing Analog and Digital Artifacts.} 
First, through generative AI, artifacts made by children and parents with various physical materials could be merged into one scene: \qt{once I created an artifact, and P4 crafted another, it allowed us to combine both creations into a unified image (P3)}. The therapeutic meaning of AI lies in its ability to \qt{harmonize the style of images. For instance, we can combine crayons and colored pencils to create together, achieving a seamless integration~(T2)}.
AI images were never intended as end products in our sessions. Therapists pinpointed that AI could \qt{shuttle}back and forth between the physical and the digital world. Physical artifacts can be blended into the digital world through AI generation; meanwhile, AI-generated images were also used as materials in physical artmaking: \qt{The AI-generated images were not necessarily final outcomes, and they could be brought to the 3D physical world for a new round of creating (T4)}. Thus, it could foster the continuity of their human creations: \qt{children can witness that their artworks have different dimensions. AI can make these artworks reused and then recreated repeatedly. I personally think it is very attractive~(T2)}.


\textbf{Generating Symbols that Can Permeate Life and Stories.} 
First, as presented earlier, AI helped children integrate personal relevance into their own stories, such as their family members and toys. 
For instance, as believed by P17, the toys brought into the story by generative AI, would increase children's connection with the story: \qt{he loves toys very much. When these toys were integrated into his stories, he might perceive them differently [...]}. 
Children might use personal toys to represent their inner world and external experiences symbolically~\cite{harvey1990dynamic}. 
Second, generative AI helped children construct self-symbols in their stories: \qt{she has immersed herself in this role because [...] she relates her current state to the character of the little rabbit~(P1)}. The therapist explained that self-symbols can help children and parents observe and reflect on their inner qualities: \qt{P1 should utilize the `little rabbit' in new stories to probe the child's life experiences or create new challenges in the stories to amplify her adaptability and other strengths~(T1)}.

On the other hand, the therapists pinpointed that the physical embodiment of AI-generated symbols, such as paper cutouts or character tokens, might serve as transitional objects. Transitional objects are typically considered as an item a child is strongly attached to in life~\cite{winnicott1951transitional}. T4 clarified the definition of transitional objects in the context of expressive arts therapy: \qt{when clients leave the therapy studio, they can take their creations, which helps them reflect back on their experience. It bridges the gap between the therapeutic environment and the real world, enhancing the transfer of therapeutic benefits and personal growth into their everyday lives}. As an example from our participants, P17 reported the character tokens became part of P18's play routines: \qt{he used those standees as his toys and played along with other toys}. This suggests that materialized AI generations have the potential to be transitional objects, which are believed to offer a sense of comfort, security, and emotional support to a child~\cite{winnicott1951transitional}.

\subsubsection{\textbf{AI as Co-creator.}} We found that generative AI, as agential materials, were not only used by families but also ``collaborated'' with them, as surfaced by two aspects:


\textbf{Family-AI Mutual Interpretations.} 
First, generative AI, by nature, interprets human input (prompts) to generate outcomes. P3 was amazed by how AI had interpreted some highly abstract inputs: \qt{P2 made a pile of mud and put it there, and AI could even create specific characters, which was quite impressive.}
Reversely, we found how AI generations served as inputs to families, prompting their creative interpretations.
Commenting on P16's case, T4 pointed out how a meaning-making process emerged in such interpretation: \qt{what fascinates me is that he not only selected the AI-image but also interpreted its unexpected part as ``jeweled bananas''. He was deeply involved in meaning-making [...] I'd consider that moment to be highly significant.} 
The families' creative interpretation of AI was incorporated into their artmaking process: \qt{I may be able to re-create stories directly based on the AI-generated images, which was quite interesting [...] For example, I didn't make trees initially, but when I saw the trees generated by AI, I was inspired to integrate this idea~(P3).} Such mutual interpretations reflect a back-and-forth conversation between the family and the generative AI tool to collaboratively shape the creative process and outcomes. 

\textbf{Family-AI Compromise and Authorship.} 
Interestingly, when the AI-generated images did not appear to align with the family-created input, most children would work with the AI's outcomes. For example, P17 explained that: \qt{initially, we probably had a preconceived notion, but when AI-generated images deviated from our expectations, we found this variation to be acceptable. 
}. The compromise might mean that children can embrace the imperfections of AI-generated images: \qt{when it comes to children, one point we may need to pay attention to is why the children think the artifacts are not perfect, and then how the children can accept this imperfection and still live with the artifacts. I think this is a relatively subtle part~(T2)}. Such compromise might foster a sense of acceptance and self-compassion~\cite{dreikurs1959courage}.

Human-AI authorship is crucial within the context of expressive arts therapy. T4 employed the connection between the creator and the creation in expressive arts therapy to stress its significance: \qt{I find myself intrigued by the dynamics of authorship between the creator and the AI-generated images. In expressive arts therapy, a client creates a drawing, and regardless of its aesthetic appeal, they often say, `this is my artwork'. However, when AI generates an image that they find beautiful, will they similarly claim authorship with confidence, `this is my creation,' when sharing it with others?}. In our study, many parents mentioned that children thought they made the story themselves: \qt{she created this story herself, and she also came up with this character~(P1)}. P5 explained that providing more choices can help establish human-AI authorship: \qt{AI provided several images and allowed them to choose one that closely mirrored their own creation. This may be considered as respecting their ideas}.

%% file: texfiles/7-discussion.tex
\section{Discussion}

\subsection{Generalizing Desirable Interaction Qualities for Children, Parents, and Therapists in Therapeutic Storymaking (\textbf{RQ3})}
Few HCI studies focused on storymaking-based expressive arts therapy in the context of family. Therefore, one of our research objectives is to contextually understand how to support families and therapists in this context. Based on our empirical findings, we discuss and generalize the desirable interaction qualities ~\cite{hook2012strong,wakkary2016unselfconscious}, to aid future research and practice in similar domains~\cite{hook2012strong,wakkary2016unselfconscious}. Summary of the desirable interaction qualities can be found in APPENDIX.

\subsubsection{\textbf{Desirable Qualities for Children}}
Four interaction qualities can be distilled to support children:

\textbf{Diverse Materials for Creative Exploration and Expression.}
As related to the Kinetic/Sensory level of ETC, offering diverse materials can benefit children's self-regulation and motor skills: e.g., cutting out paper or manipulating clay in \autoref{fig:etc}~\circled{1}~\circled{2}). ETC argues for the significant therapeutic value of (analog) material interactions~\cite{hinz2019expressive}, which cannot be easily replaced by digital interaction. Embodied interaction with rich materials could also enrich children's non-verbal expression. Materials serve as a medium for emotional expression for children who may not express their feelings through verbal communication~\cite{wikstrom2005communicating}. Most importantly, our data showed combining analog materials with AI materials promised huge creative empowerment for children. The rich artifacts in \autoref{fig:etc} concretize how AI-infused multimateriality facilitated their expressing of feelings and self-actualization~\cite{manheim1998relationship}, which are the core therapeutic values for children.

\textbf{Personal Relevance.}
Personal relevance is an essential quality for children's engagement. We found that many children tended to identify themselves or their family members as story characters in their family stories~(e.g, \autoref{fig:etc}~\circled{11}). These characters are the symbolic instruments for children to project inner emotional states and externalize inner conflicts through stories~\cite{jennings2005embodiment}. Children's personal toys were also integrated into their stories, potentially serving as comforting and empowering tools for self-expression and their emotional connection to their stories~\cite{kottman2014play}.

\textbf{Sense of Accomplishment.} 
The sense of accomplishment came from artifacts co-created with AI which exceeded children's expectations. 
The process of crafting physical artifacts for AI input and choosing from AI-generated variations encouraged children to assert ownership, feeling as if the creations were their own rather than the AI's.
Moreover, the sense of accomplishment also came from that children could share their own storybooks with family members or their peers~(e.g., \autoref{fig:teaser}~\circled{d}). Such family sharing enhances the sense of family togetherness and unity, benefiting children's growth and learning~\cite{christensen2019together}.

\textbf{Renarration and Recreation beyond the Therapeutic Sessions.} 
In our study, many children took the AI-generated materials home and continued making new stories at home~(e.g, \autoref{fig:finding1}~\circled{24}). They utilized not only the paper cutouts but also physical materials from daily life~(e.g., \autoref{fig:finding1}~\circled{23}). This suggests a valuable quality of our activities that helped break down the boundaries between the therapeutic environment and real life. This encourages continued creative self-explorations and self-expression to take place after the therapeutic interventions~\cite{panjwani2017constructing}.

\subsubsection{\textbf{Desirable Qualities for Parents}}
 Three qualities were discovered to support parents, which can be further amplified by future HCI systems:

\textbf{Discovering New Aspects of Children.}
First, many parents appreciated that the artifacts and process of their family storymaking helped uncover children's demands for parents. Through interacting with or through story characters symbolizing the family members, parents could gain insights into how children view these relationships and the dynamics at play~\cite{kwiatkowska1967family}.

Also, parents discovered children's less noticeable life experiences through connecting stories and life~(e.g, P10's kindergarten experiences, see \autoref{fig:finding1}~\circled{7}). Storytelling is a powerful tool to help parents make sense of children's life experiences and identities~\cite{rubegni2019detecting}.
Finally, parents uncovered their children's strengths through making stories. For example, P3 was surprised to recognize the heroic valiant characteristic within P4~(see \autoref{fig:etc}~\circled{15}). 
As T1 suggests, parents could thoughtfully create new stories with children to continue reinforcing these observed strengths. 

\textbf{Reflection in Therapeutic Storymaking.} We found that during the storymaking, some parents meaningfully reflected on their roles and their family interactions in the past. Two parents became aware of their dominant supervisory role during the activities. Such reflection can improve their family roles and family flexibility~\cite{olson2000circumplex}. Additionally, some parents realized they tended to focus excessively on the intricacy of the storylines crafted by children. After discussions with the therapist, a few parents noted they should learn to appreciate the rawness or genuineness of the stories, which align with their children's age and lived experiences.

\textbf{Storymaking as Family Meaning Co-constructing.} Several parents mentioned that the storybooks co-created with their children could become part of their family memory archive~(see \autoref{fig:finding1}~\circled{18}).
As T1 envisioned, if storymaking becomes a family ritual over an extended period, parents would be able to~\qt{witness the growth of children~(T1)} over time, and these stories can \qt{reflect what the children care about at the moment~(T1)} by compiling a series of family stories just like a family photo album.
In our study, several families indeed used paper cutouts or character tokens of AI-generations at home for making new stories. This suggests the potential value of promoting family interaction~\cite{christensen2019together} and the sense of togetherness~\cite{wajskop2015dramatic} when storymaking can become a family ritual.

\subsubsection{\textbf{Desirable Qualities for Therapists}}
Three interaction qualities are desirable when supporting therapists:

\textbf{Observing the Strengths and Challenges of Children and Families.} Therapists constantly try to observe the strengths of the children via storymaking activities, and utilize specific techniques to reveal these strengths to both children themselves and their parents while steering the process.
According to positive psychotherapy theory~\cite{rashid2015positive}, such a \textit{strength-based} approach can enable therapists to identify and capitalize on children's strengths, granting resources for children to face the challenges in life.
Therapists also try to gain insights into family interaction dynamics and their challenges in communication and relationship through the process and outcome artifacts of the family's joint storymaking~\cite{kerr2011family}. 
For example, P13 explicitly projected himself, his mother, and his father to a purple fish, a seahorse, and a jellyfish. His story offered T1 insights into their family dynamics:~\qt{when the purple fish felt cold, he went to the seahorse. The jellyfish does not have a home [...] He simply drifts along~(T1)}. T1 thereby confirms the child's strong attachment to the mother and the sense of disconnection of the father due to the father's limited availability for family. Also, T2 identified family interaction issues by observing a mother's frowning responses to the artifacts her child presented.

\textbf{Moderating Family Communication through Storymaking.} 
First, therapists often need to facilitate communication through role-playing. During the activity, T1 encouraged the family members to name the characters and play the story characters created by themselves to express their inner thoughts. For example, T1 facilitated P1 and P2 to take on their created characters to communicate in the playground scene~(see \autoref{fig:finding1}~\circled{16}). Children might express their thoughts more comfortably through role-playing, leading to open and honest conversations~\cite{ahmadpour2023understanding,matthews2014taking}. 
Further, therapists commonly need to initiate perspective-taking plays to enhance mutual understanding among family members. T1 encouraged the child to take on the role of the elephant mother created by his mother to act out what the angry elephant mother would say to the baby elephant using AI-image cutouts during the activities. Facilitating perspective-taking can help families overcome egocentric tendencies, promote mutual understanding, and foster empathy~\cite{corsini2017role}.

\textbf{Offering Families a Playful and Safe Space.} 
Therapists have a substantial need to afford a playful space for family storymaking. Family storymaking is \qt{a process of play~(T3)}, and play \qt{allows us to better release our personality so that we can be ourselves~(T3)}, which is similar to previous research findings~\cite{hubbard2021child}.
We found that the therapist intended to offer diverse playful creative tasks from crafting to dramatic play. Children enjoyed playing and performing using character tokens, personal toys, or body movement~(see \autoref{fig:finding1}~\circled{15} and \circled{17}). 
Also, T1 could help parents return to a child-like mindset by providing diverse analog materials~(see \autoref{fig:teaser}~\circled{e}). For instance, P5 noted that despite she bought clay for her child before, she had never used it herself; and during the session, she was impressed by how relaxed she felt when interacting with it. 
And her relaxation also positively influenced her child's emotional state.
Further, therapists value the importance of affording a safe space where participants feel comfortable while engaging in the creative process and expressing their thoughts~\cite{lazar2018making}. In our study, a safe space was interpreted as \qt{a loving environment~(P3)}, where a child can feel that \qt{everyone accepted him and encouraged him to express himself freely without concerns~(P3)}. 

\subsection{Design Implications (\textbf{RQ3})}
Besides the desirable interaction qualities, we also formulate a set of design implications with the purpose of inspiring future HCI systems targeting AI-infused, multimaterial storymaking for family expressive arts therapy (\textbf{RQ3}).


\textbf{Implication 1: Supporting Materialization of Generative AI in Family-AI Engagement.} 
The diverse range of examples gathered through the four levels of ETC illustrates that children combined diverse analog materials and AI materials to craft stories. Prior research has shown that AI might wield significant influence at higher-level components, such as affective ~\cite{wang2023reprompt} or cognitive components~\cite{han2023design}. 
However, our study, applying Expressive Therapies Continuum (ETC), revealed new insights into family-AI interaction design, applicable beyond therapy contexts.
For instance, it has been revealed that by enabling \textit{physicalization} or \textit{materialization} of AI-generated contents, we could enrich the sensorial and kinesthetic qualities of family-AI interactions, which could not only aid children's motor-sensory skills but also induce mental soothing effects~\cite{lusebrink1990imagery} and positive parental involvement~\cite{yu2021parental}. 
Moreover, looking through the creative dimension of ETC, we found that such materialization of AI generated contents could boost families' creative potential by allowing them to combine analog materials with AI materials in physical artmaking activities. 
These insights could be leveraged beyond the expressive therapy context to a broader range of applications which means to support comforting and enjoyable family-AI co-creative tasks, through tangible, embodied, or mixed reality interfaces. 

Utilizing a therapeutic perspective, we also uncovered for the HCI community how tangible character tokens created in family-AI storymaking can act as \textit{transitional objects}~\cite{winnicott1951transitional}.
These objects can offer children continued companionship beyond the creative activity, holding the potential to aid children's self-empowerment journey in everyday life.
This could inspire future child-AI systems beyond the therapy contexts to leverage the developmental benefits of child-AI co-created characters as both digital transitional objects~\cite{koles2016avatars}, and physical transitional objects, for example, embodied by customized tangible toys, robotic systems~\cite{luria2020destruction}, or wearable companions~\cite{ozcan2016transitional}, in order to more ubiquitously grant children with emotional and developmental support in daily surroundings~\cite{luria2020destruction}.

\textbf{Implication 2: Supporting Constructing Self-symbols for, and with Children.} 
In our study, we emphasized the paramount importance of crafting story characters using AI that are intricately connected to the concept of self-symbols~(e.g, \autoref{fig:etc}~\circled{11}). 
The AI-generated images of characters were based upon the children's artifacts or toys, and they were strongly appreciated for enabling high-quality, coherent characters that were seamlessly blended into diverse story scenes with a harmonious presence. These characters were found to be used as children's self-symbols or projections of family members, which can help children authentically reflect in the stories on the qualities inherent to their identities~\cite{hinz2019expressive}. 
We thereby suggested that future research could intentionally integrate story-relevant avatars created through the generative AI to probe children's mental projection~\cite{zarei2021towards,zarei2020investigating} in daily lives, which can help them to be better understood by their parents and therapists. 
These avatars might be applied in educational games for children to increase personal relevance or overcome mental challenges in the learning process ~\cite{lee2020using,koles2016avatars}. 
Moreover, the avatar can be designed to serve as a conversational agent, providing companionship and fostering growth alongside children~\cite{xu2022elinor,xu2023rosita}.

\textbf{Implication 3: Supporting Moderation of Family Communication via Storymaking.} 
We found that the therapist frequently need to employ role-playing and perspective-taking to moderate the communication within families. Engaging in role-playing and perspective-taking can help parents and children comfortably and empathically communicate via story characters, strengthening their understanding of each other's thoughts and feelings~\cite{shuttleworth1980drama}. The future HCI design could enable interactive therapeutic systems to leverage and augment role-playing and perspective-taking through technology-mediated interactions. To encourage therapeutic role-playing~\cite{matthews2014taking}, the system can engage with parents and children in various role-playing scenarios using human-AI co-created story characters. To facilitate perspective-taking, we suggest that the system encourages parents to actively take on the role of the help-seeker while the children can take on the role of problem-solver or caretaker, fostering children's inner growth and parents' empathy with children. Moreover, the system can introduce different conflict points within the story plots where parents and children need to switch roles of characters and make collective efforts. For instance, think of a collaborative VR game combining storymaking and escape room, requiring parents and children to co-create stories and play each other's roles to proceed. 

\textbf{Implication 4: Supporting Family Meaning Co-constructing and Family Memory Archiving via Storymaking.} 
In our findings, we found that the storybooks co-created by the families could become a meaningful part of their family memory archive. 
This echoes previous research which has demonstrated that sharing stories as well as creating stories can help to co-construct family meaning~\cite{jones2018co,wallbaum2018supporting}. 
In the future, an AI-infused multimaterial storymaking system can be designed to intentionally support parents and children recording, renarrating and reliving multidimensional life experiences, such as kindergarten activities, family traveling, or other recreational activities. Also, future work could broadly explore meaningful forms of interactions or outcomes artifacts for family memory archiving and revisiting, such as stop-motion animation, short videos, mobile wallpapers, 3D photos, mementos, or interactive storybooks. Further, a built-in online community component can facilitate the storage and exchange of family memories related to the collaborative storymaking process, fostering connections both within and across families and affording a platform for the ongoing evolution of their shared narratives. 

\subsection{Limitations and Future Opportunities}
There are several limitations in our study. First, generative AI might not always produce preferred outcomes due to inherent uncertainties in AI. To counter this, we afforded multiple cycles of generation and regeneration, allowing families to iteratively explore different outcomes. This involved adjusting their input images (photos of their artifacts) and verbalization (used for text prompts) until satisfactory results were achieved, sometimes requiring several iterations.
While this highlights a limitation in current generative AI technology, it also presents a future opportunity to enhance user experience through HCI approaches. For example, developing a more intuitive input interface could facilitate easier uploading and pre-processing of image inputs and modification of text prompts. A well-designed interface with structured prompting questions could potentially reduce the number of iterations needed to achieve desired results.
Furthermore, this limitation can be reframed as a design opportunity to encourage family communication and children's expressive skills. The need for both image and verbal inputs, and the family's discussion around AI-generated outcomes, can be intentionally designed to foster children's verbal expression and family communication. This may not only aid in expressive arts therapy but also support overall child development and strengthen family relationships.

Another important aspect regards the safe use of generative AI. 
We selected Midjourney from various available alternatives because it has a robust mechanism for preventing the output of inappropriate, misleading, or unsuitable content~\cite{midGuidelines}. 
While we did not encounter unsuitable images, there were occasional minor issues with counterfactual or unconventional elements in the AI-generated images. For instance, a cartoon snake might be depicted with feet.
While this issue is common in current generative AI technologies,
future systems could consider incorporating additional prompts in the back end to avoid specific types of errors, 
or, intentionally focusing on generating less realistic, more cartoonish styles to avoid discomfort or uncanny effects from photo-realistic images with errors.
Interestingly, our study also found that these counterfactual elements can spark creativity in children. 
For instance, a child used an AI-generated image of a banana with unexpected droplet-like features to invent a ``diamond banana'' concept in his story. 
This suggests that the unpredictability of AI can sometimes be a source of creative inspiration.
However, it is crucial to recommend that future research collaborate closely with parents, therapists, children, and other stakeholders to ensure the safe use of generative AI with children and develop design principles for age-appropriate generative AI interface, minimizing potential risks while maximizing creative opportunities~\cite{wang2022informing}.

Here we reflect on how our study might contribute to a few ongoing conversations in HCI.
A key issue in HCI research is enabling better family communication and relationships through emerging technologies. 
Our study highlighted how family expressive arts therapy offers novel avenues for creating family-oriented technologies. 
This includes integrating multiple forms of expressive activities in HCI designs to foster playful, therapeutic interactions between family members.
Techniques widely used in family therapy, such as role-playing~\cite{matthews2014taking} or perspective-taking~\cite{ohtaka2019perspective} among family members, could inspire future technology designs for facilitating family engagement and mutual understanding. 
Our study also concretized the value of family storymaking not just for children's learning but for preserving family memories and jointly constructing family meaning over time~\cite{jones2018co}.
Additionally, we surfaced the exciting possibilities of combining generative AI with embodied or physical interaction experiences.
Lastly, along with the emerging discussion about family engagement with AI, our study surfaced a relevant and concrete understanding of the needs and experiences of families using AI for creative expression. This contributes valuable, practical design insights for future initiatives.

%% file: texfiles/8-conclusion.tex
\section{Conclusion}
In this study, we set out to empirically explore how to utilize image-based generative AI as expressive materials and how to support the needs of children, parents and therapists in the context of family expressive arts therapy. In doing so, we collaborated with a professional therapist to conduct five-week expressive arts therapy sessions involving 18 participants from 7 different family groups. Our findings highlighted that
emerging outcomes and process of storymaking-based expressive arts therapy combining generative AI and analog materials. Through the lens of ETC, we characterize the therapeutic meanings of generative AI as expressive materials. Our work contributes to a nuanced understanding of desirable interaction qualities of children, parents and therapists, and provides design implications regarding supporting storymaking based expressive arts therapy.

\begin{acks}
We are grateful to all the collaborating therapists for their generous support. Our special thanks are extended to Therapist Jia Liu for her remarkable dedication and invaluable contribution to this research. Our appreciation also goes to all the participating families for their time and efforts in the participation. This work is supported by the National Social Science Fund of China (NSSFC) Art Grant: No. 22CG184.
\end{acks}